\documentclass[lettersize,journal]{IEEEtran}
\usepackage{amsmath,amsfonts}
\usepackage{algorithmic}
\usepackage{algorithm}
\usepackage{array}
\usepackage[caption=false,font=normalsize,labelfont=rm,textfont=rm]{subfig}
\usepackage{textcomp}
\usepackage{stfloats}
\usepackage{url}
\usepackage{verbatim}
\usepackage{graphicx}
\usepackage{cite}
\usepackage{caption}
\usepackage{float}
\usepackage{color}
\usepackage{pifont}
\usepackage{enumerate}
\usepackage[acronym,nonumberlist,nopostdot]{glossaries}
\usepackage{glossaries}
\usepackage{glossary-mcols}
\usepackage{nomencl}
\usepackage{setspace}

\renewcommand{\baselinestretch}{1}
\newcommand{\tabincell}[2]{\begin{tabular}{@{}#1@{}}#2\end{tabular}}
\begin{document}

\title{End-to-End Design of Polar Coded Integrated Data and Energy Networking}

\author{Jie Hu, {\em Senior Member, IEEE}, Jingwen Cui, Luping Xiang, {\em Member, IEEE}, Kun Yang, {\em Fellow, IEEE}
\thanks{This work was supported in part by MOST Major Research and Development Project (Grant No.: 2021YFB2900204), Natural Science Foundation of China (No. 62132004), UESTC Yangtze Delta Region Research Institute-Quzhou (No. 2022D031, 2023D005).}
\thanks{Jie Hu, Jingwen Cui, Luping Xiang and Kun Yang are with the School of Information
and Communication Engineering, University of Electronic Science and Technology of China, Chengdu 611731, China, email: hujie@uestc.edu.cn, 202221011120@std.uestc.edu.cn, luping.xiang@uestc.edu.cn, kyang@ieee.org. \textit{(Corresponding author: Luping Xiang.)}}
}



\maketitle

\begin{abstract}
In order to transmit data and transfer energy to the low-power Internet of Things (IoT) devices, integrated data and energy networking (IDEN) system may be harnessed. In this context, we propose a bitwise end-to-end design for polar coded IDEN systems, where the conventional encoding/decoding, modulation/demodulation, and energy harvesting (EH) modules are replaced by the neural networks (NNs). In this way, the entire system can be treated as an AutoEncoder (AE) and trained in an end-to-end manner. Hence achieving global optimization. {Additionally, we improve the common NN-based  belief propagation (BP) decoder by adding an extra hypernetwork, which generates the corresponding NN weights for the main network under different number of iterations,} thus the adaptability of the receiver architecture can be further enhanced.
Our numerical results demonstrate that our BP-based end-to-end design is superior to conventional BP-based counterparts in terms of both the BER and power transfer, but it is inferior to the successive cancellation list (SCL)-based conventional IDEN system, which may be due to the inherent performance gap between the BP and SCL decoders. 
\end{abstract}

\begin{IEEEkeywords}
Integrated data and energy networking (IDEN), wireless energy transfer (WET), polar code, deep neural network (DNN)
\end{IEEEkeywords}

\begin{table}[h]
\centering
\caption*{ACRONYMS}
\begin{tabular}{ll}
{AE} & {AutoEncoder}\\
AWGN & Additive White Gaussian Noise\\
BER & Bit Error Rate\\
{BLER} & {Block Error Rate}\\
BP & Belief Propagation\\
BNN &  Binarized Neural Network\\
CE &  Cross Entropy\\
{CRC} & {Cyclic Redundancy Check}\\
DC & Direct Current\\
DE & Density Evolution\\
DNN & Deep Neural Network\\
EH & Energy Harvester\\
GA &  Gaussian Approximation\\
IDEN & Integrated Data and Energy Networking\\
IoE & Internet of Everything\\

\end{tabular}
\end{table}
\begin{tabular}{ll}
IoT & Internet of Things\\
LLR & Logarithmic Likelihood Ratio\\
ML & Machine Learning\\
NN & Neural Network\\
OF & Objective Function\\
PAPR & Peak to Average Power Ratio\\
{PE} & {Processing Element}\\
{PPV} & {Polyanskyi-Poor-Verd$\rm{\grave{u}}$}\\
{PS} & {Power Splitting}\\
PW & Polarization Weight\\
QAM & Quadrature Amplitude Modulation\\
{QPSK} & {Quadrature Phase Shift Keying}\\
ReLU & Rectified Linear Unit\\
RNN & Recurrent Neural Network\\
RF & Radio Frequency\\
{SC} & {Successive Cancellation}\\
{SCL} & {Successive Cancellation List}\\
SGD & Stochastic Gradient Descent\\
SNR & Signal-to-Noise Ratio\\
Tanh & Hyperbolic Tangent\\
WET & Wireless Energy Transfer\\
WIT & Wireless Information Transfer\\
3GPP & Third Generation Partnership Project\\
{5G} & {Fifth Generation}\\
\end{tabular}


\section{Introduction}
\subsection{Background}
Until recently, information and energy were transmitted across the network separately. However, as a benefit of power efficient electronic devices, {combining wireless information transfer (WIT) with wireless energy transfer (WET) is worth studying in the wireless network design field, which inspired the emergence of the integrated data and energy networking (IDEN) concept \cite{hu2018integrated}}. The IDEN technique makes full use of the RF signals, enabling them to transmit information and energy simultaneously. Thus it can provide energy supply for numerous low-power devices in the Internet of Everything (IoE) and enable them to communicate with each other \cite{clerckx2022foundations1,clerckx2021wireless1,clerckx2018fundamentals}. 

Traditional designs of IDEN systems are based on communication and information theory, mathematical analysis and modeling, as well as convex optimization, supported by rigorous and reliable theoretical analysis. However, IDEN systems have numerous nonlinear components, such as energy harvesters (EH), which are difficult to model accurately. Therefore, researchers have turned their attention to machine learning (ML) techniques to seek new system designs, since the learning based approaches may not require precise modeling and mathematical analysis. For instance, Varasteh \textit{et al.} \cite{varasteh2020learning} proposed a deep neural network (DNN)-based IDEN system, where the transmitter and the receiver are interpreted as an AE, demonstrating the great potential of the learning based approaches for IDEN systems. {We extended the DNN-based IDEN architecture of \cite{xiang2023polar} to a polar-coded IDEN system,} obtaining superior performance over the traditional counterparts in terms of both the WIT and WET.  

\subsection{Related work}
Since both information and energy are transmitted simultaneously in the IDEN system, there is a compromise between WIT and WET, which means that a better communication performance is typically accompanied by a degradation in energy transfer performance, and vice versa. Therefore, how to balance WIT and WET to strike a trade-off becomes the main challenge in the IDEN field. Meanwhile, the low energy transfer efficiency is also an obstacle in WET \cite{clerckx2018fundamentals}. To alleviate these problems, substantial advances have been made in signal processing, system designs and EH model designs \cite{fouladgar2014constrained,tandon2016subblock,basu2019polar,dabirnia2016code,kim2021wireless,zeng2017communications,clerckx2016waveform,ku2017joint,collado2014optimal,huang2017waveform,varasteh2020oncapacity,clerckx2018beneficial0,clerckx2017wireless11,varasteh2019learning2}.

Physical layer techniques of coding, modulation, waveform design and beamforming have been developed. In terms of coding, the constrained run-length limited codes of \cite{fouladgar2014constrained} and the sub-block energy-constrained codes of \cite{tandon2016subblock} both considered trade-offs between having sufficient energy and high information transfer rates.
Basu \textit{et al.} \cite{basu2019polar} utilized polar codes for improving the capacity vs. energy performance by either concatenating nonlinear mappings with linear polar codes or using randomized rounding for polar codes. Dabirnia \textit{et al.} \cite{dabirnia2016code} adopted the connection of nonlinear trellis codes and low density parity check codes in IDEN, showing a superior performance, which is only about $0.8$dB away from the information theoretic limits. In the modulation field, Kim \textit{et al.} \cite{kim2021wireless} designed a pulse position modulation scheme for IDEN systems, enabling the transmission of signals having high peak to average power ratio (PAPR) and consequently maximizing both  the communication and energy transfer performance.
In addition, the waveform designs \cite{zeng2017communications,collado2014optimal,clerckx2016waveform,huang2017waveform,ku2017joint} are closely related to the energy conversion efficiency from RF to direct current (DC) power. Collado \textit{et al.} \cite{collado2014optimal} demonstrated that signals having high PAPR were capable of providing efficiency improvements. A novel multisine waveform \cite{clerckx2016waveform} was designed for WET, which exhibited significant DC power gains. In \cite{huang2017waveform}, a pair of limited feedback-based waveform strategies were adopted for WET and harvested more energy. Moreover, Ku \textit{et al.} \cite{ku2017joint} conceived a joint waveform as well as beamforming design and improved the energy transfer efficiency. 
In addition to efficient signal processing techniques, the nonlinear modeling of EH also plays an indispensable role in improving the energy conversion efficiency. A large number of contributions \cite{clerckx2016waveform,varasteh2020oncapacity,clerckx2018beneficial0,clerckx2017wireless11} have proved that linear EH models are inaccurate due to the nonlinear characteristic of the diodes in the rectifier, while nonlinear modeling is more consistent with practical EH models. Varasteh \textit{et al.} \cite{varasteh2019learning2} proposed a pair of nonlinear EH models for large and small input power regimes, respectively, in order to improve the modulation format of IDEN systems. 

The above mentioned solutions to the trade-off between information and energy as well as the efficiency improvement of energy transfer in the IDEN system are based on the traditional models and convex optimization methods. Although these methods are effective and implementable, they still have limitations in terms of analyzing and formulating the nonlinear models of components (e.g., rectennas, nonideal filters and diodes in the EH) by mathematical methods, which leads to a bottleneck in IDEN designs. As a remedy, researchers conceived ML techniques to assist the associated signal processing and system modeling, expecting a ``learning'' based method to tackle these challenges. 
In the signal processing field, a DNN-based energy beamforming scheme was proposed for solving the resource allocation problem of the WET and WIT in \cite{9583222Hameed}. Kang \textit{et al.} \cite{8469031Kang} applied deep learning to extract the channel state information and demonstrated superior performance. In terms of the EH models, considering the problem that the nonlinear components like diodes are hard to analyse accurately by mathematical formulas, deep learning became an instrumental technique of alleviating this problem, since it can avoid modeling by mathematical analysis. {In \cite{varasteh2020learning}, DNNs were utilized to model EH and to train it with the aid of a nonlinear regression algorithm over large amounts of data for attaining high accuracy. Their DNN-based EH was embedded in an end-to-end IDEN design and performed energy harvesting. 
Shanin \textit{et al.} \cite{shanin2020rate} modelled the memory effect of the non-linear EH by a Markov reward chain, where the output voltage levels and the harvested power were treated as states and rewards of the chain, respectively. The DNNs were utilized to simulate the Markov chain to fit the specific non-linear characteristic of actual EH circuits. }

\begin{figure*}
  \centering
  \includegraphics[width=6.3in]{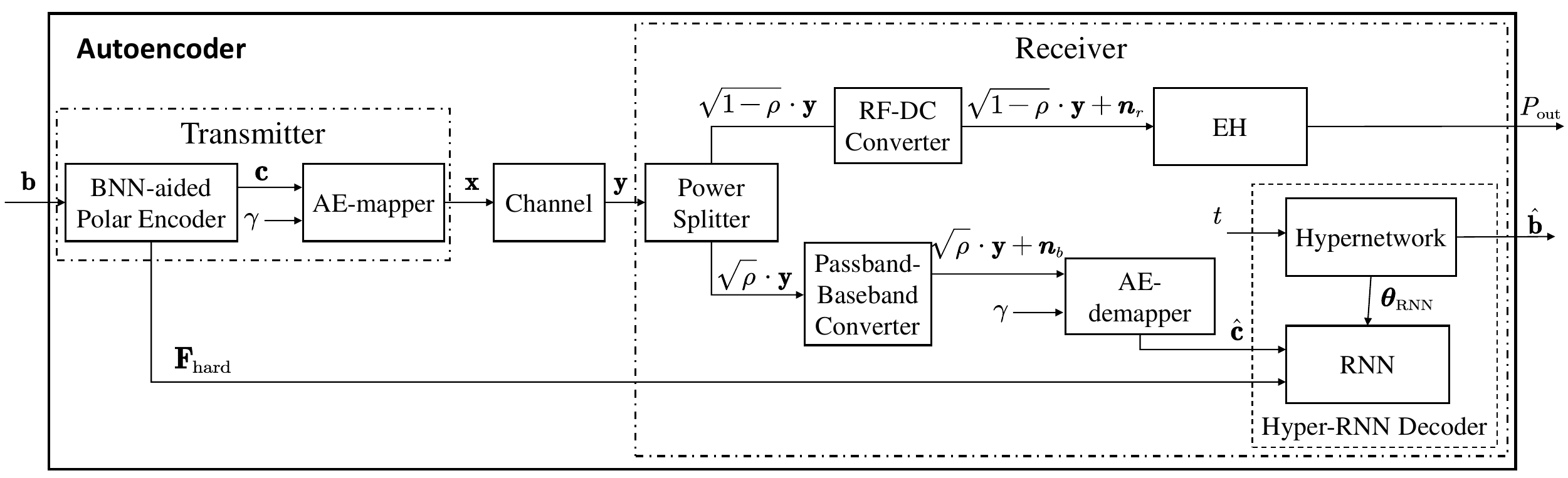}\\
  \caption{An end-to-end polar-coded IDEN system.}\label{fig.systemmodel}
\end{figure*}

Current research on coding techniques for IDEN systems is limited, but the results of \cite{basu2019polar} demonstrated the potential of polar codes in this novel context. Polar codes were originally  proposed by Arikan \cite{arikan2009channel} and have substantially evolved since then. 
Numerous researches focused on the learning-based encoding and decoding algorithms of polar codes \cite{9097207Xu,8863406Liu,8109997,teng2019low,ebada2019deep}. Liu \textit{et al.} \cite{8863406Liu} proposed a deep learning based cyclic redundancy check (CRC)-error-correcting aided SCL decoder by implementing a long short-term
memory network. The authors of \cite{8109997} and \cite{teng2019low} conceived neural network (NN)-based BP aided decoders by exploiting the structural similarity between the polar code's factor graph and that of the neural networks. In addition, Ebada \textit{et al.} \cite{ebada2019deep} designed a so-called binarized neural network (BNN) to construct novel polar codes and showed superior performance compared to the conventional 5G polar codes \cite{8962344Bioglio}. However, the effect of polar codes on the WET performance has hitherto been ignored in the study of IDEN.

\begin{table}[t]
\centering
\footnotesize
\caption{Contrasting Our Contributions To The State-Of-Art}
\begin{tabular}{l|c|c|c|c|c } 
 \hline
 Contributions & \textbf{this work} & \cite{xiang2023polar} & \cite{basu2019polar} & \cite{ebada2019deep} & \cite{varasteh2020learning} \\ \hline\hline
 Polar coded IDEN & \ding{52} & \ding{51} & \ding{51} &  &  \\ 
 \hline
 NN-based polar encoder & \ding{52} &  &  & \ding{51} &  \\ 
 \hline
 Adaptive polar decoder & \ding{52} &  &  &  &   \\
 \hline
 Joint optimization & \ding{52} & \ding{51} &  &  & \ding{51} \\
 \hline
\tabincell{l}{Fully DNN-based coded\\ IDEN }  & \ding{52} &  &  &  &  \\
 \hline
 \tabincell{l}{Adaptive power splitting \\principle } & \ding{52} &  &  &  &  \\
 \hline
\end{tabular}
\label{tab.contrast}
\end{table} 

While learning-based physical technologies have been investigated in the context of IDEN systems \cite{8469031Kang,9583222Hameed,varasteh2019learning2}, they still have some drawbacks. For instance, the NNs are only employed to replace specific functions, e.g., beamforming or modulation, which only achieves local optimization instead of global optimization and may also affect the overall system performance. In addition, the current research of deep learning based coding techniques for IDEN is still in its infancy, although the coding scheme constitutes an important component of the IDEN system, yet the beneficial impact of polar codes on WET performance in IDEN system has been overlooked.
Exploring the characteristics of polar codes reveals that incorporating the energy requirements into the associated polar code construction is capable of improving the WET performance and balance the information and energy networking in IDEN systems. However, the frozen set selection of conventional polar code construction is generally based on the channel reliability, which is typically calculated by mathematical algorithms, such as the Bhattacharyya Parameter and Density Evolution (DE) of \cite{mori2009performance}, which tends to impose both excessive computational complexity and coding latency. The novel polar encoding algorithm proposed for the 3GPP 5G system \cite{3rd2021technical} is universal, but it may not be compatible with IDEN systems, since the variable fading channels and energy transfer requirements are bound to affect the system's choice of transmission channels, while the fixed frozen bit strategy adopted by \cite{3rd2021technical} fails to meet the above dynamic requirements. Hence it would lead to performance degradation in IDEN.
Moreover, although Xu \textit{et al.} \cite{8109997} have achieved a satisfactory BER performance at a low complexity, their pure DNN-based decoder can not achieve its expected performance, when the training iterations and test iterations are different, which reflects its limited adaptability. Therefore, the pure DNN-based decoding algorithm needs further improvements to cope with the above requirements. 


\subsection{New Contributions}
Against this background, we conceive an end-to-end polar-coded IDEN system aiming for enhancing both the WET as well as WIT performance. {We adopt the AE architecture, which considers the transmitter and receiver as a whole, enabling joint training of all modules. Therefore, the entire IDEN system is treated as an AE and trained in an end-to-end manner to achieve global optimization.} {Polar codes are considered mainly due to the capacity-achieving property. They have been adopted by the 3GPP 5G NR standard as the official coding scheme for the control channels. Therefore, its advantage in the communication field inspires us to explore its application also in the IDEN field. The superiority of polar codes in communication means that more resources can be used for satisfying the energy requirements. }
{The EH model proposed in \cite{varasteh2020learning} is adopted in our system, because our requirement for the EH in the IDEN system is to directly obtain the mapping between the received signal and the harvested energy, without caring about the internal process of EH. The model in \cite{varasteh2020learning} is the most suitable one, since it directly learns the input power/output power relationship by applying nonlinear regression, while the model in \cite{shanin2020rate} is based on the Markov reward chain, which involves state transitions and reward mechanisms, thus it is more complex and not applicable to our system.}
Our main contributions are boldly and explicitly contrasted to the existing literature in Table \ref{tab.contrast} and they are summarized in more detail as follows:
\begin{itemize}
  \item We propose a BNN-aided polar encoder, where the frozen/information bit positions are selected by a BNN. The frozen/information bit positions are interpreted as a trainable vector, which can adjust itself to adapt to the time-variant channels and satisfy the energy requirements during the training process. By employing the learning-aided frozen set selection strategy, the IDEN system becomes capable of achieving both WIT and WET performance improvements.
  
  \item We conceive a hyper-RNN polar decoder, which extends the pure DNN-based BP decoder to a pair of neural networks exhibiting flexible adaptability. The hypernetworks take the number of decoding iterations $t$ as their input, and generate the corresponding weights as a function of the number of iterations $t$ for attaining near-optimal decoding performance. Therefore, the hyper-RNN decoder can be specifically trained for particular number of iterations to enhance its adaptability, which perfectly solves the problem of weak adaptability when using a pure DNN-adied decoder.

  \item We propose a fully DNN-based coded IDEN system, where the polar encoder, the modulator, the demodulator, the polar decoder and the EH are all based on DNNs, which greatly exploits the potential of DNNs for achieving the optimal IDEN performance. Additionally, we conceive an adaptive power splitter by interpreting the power splitting factor as a trainable variable, thus the power splitter can better satisfy the practical data vs. energy requirements by  adjusting the factor.

  \item {Our numerical results demonstrate that the polar coded end-to-end IDEN system relying on our BNN-aided polar encoder, DNN-based modulator/demodulator and hyper-RNN decoder improves both the WIT and WET performance compared to the traditional BP-based IDEN system explicitly. Approximately $0.005$ mW power improvement is attained at BER $=1.6\times10^{-4}$ at the transmit power of $3$ dBm for transmission over AWGN channels. This is further improved to $0.028$ mW at BER $=3.9\times10^{-2}$ at the transmit power of $0$ dBm over Rayleigh channels.}
\end{itemize}

The rest of this paper is organised as follows. Section II describes our system model, while Section III introduces the details of the optimization problem and the corresponding solution. After presenting the simulation results in Section IV, we finally conclude in Section V.

\section{System Model}
In this section, the structure of our polar coded IDEN system seen in Fig.\ref{fig.systemmodel} is presented. Then, the BNN-aided polar encoder and the AE-mapper of the transmitter, as well as the power splitter, the RF-DC converter, the passband-baseband converter, the EH, the AE-demapper and the hyper-RNN decoder of the receiver are described in detail.
\subsection{Transmitter}
\subsubsection{BNN-aided Polar Encoder}\label{sec.polar encoder}
{The encoding process of an $(N,K)$ polar code is based on the channel polarization, through which each of the $N$ sub-channels will have a different reliability. A higher reliability means a higher probability that a channel transmits a single bit and that it is correctly decoded. If $N$ is large enough, the synthesized channels tend to two extremes: part of the channels become noisy channels with low reliability, while the other part hosts noise-free channels with high reliability \cite{8962344Bioglio}. The $K$ information bits to be transmitted will be assigned to $K$ sub-channels with the highest reliability among the total of $N$ sub-channels, while the remaining sub-channels transmit the zero-valued frozen bits.}

Generally, the polar code construction relies on selecting the frozen bit positions, which depends on the channel reliability \cite{babar2019polar,Egilmez8936409}. Common polar construction algorithms such as Gaussian Approximation (GA) and DE are widely used in communication systems. However, considering an IDEN scenario, where the WET requirement also affects the transmitter's choice of the transmission sub-channels, the common construction algorithms exhibit weaknesses in determining the frozen/information bit positions, since they cannot take the WET impact into account when calculating the channel reliability. {In order to solve the polar code construction problem of IDEN systems, we follow the spirit of \cite{ebada2019deep}, and a BNN is harnessed for the frozen set selection in an IDEN scenario, which can adjust the probability of the sub-channel being information bit position according to both the WET and WIT feedback}, and make the encoder more compatible with the IDEN system.

\begin{figure}[t]
  \centering
  \includegraphics[width=2.8in]{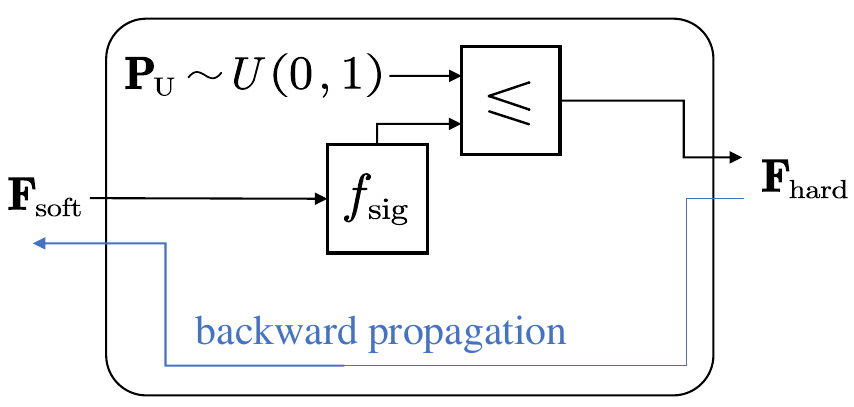}\\
  \caption{The architecture of BNN.}\label{fig.BNN}
  \vspace{-0.4 cm}
\end{figure}

The basic structure of BNN is demonstrated in Fig.\ref{fig.BNN}. It can be observed that the probabilities of being information bit position of all sub-channels are treated as a vector variable $\mathbf{F}_{\textrm{soft}}$ and it constitutes the input of the BNN. Consider an $(N,K)$ polar code as an example, in its forward propagation, the input $\mathbf{F}_{\textrm{soft}}$ of length $N$ is processed by the Sigmoid function $f_{\textrm{sig}}$ and it is mapped into a probability vector $\mathbf{F}_{\textrm{pro}} = f_{\textrm{sig}}(\mathbf{F}_{\textrm{soft}})$. The vector $\mathbf{F}_{\textrm{pro}}$ represents the probabilities of total $N$ sub-channels representing the information bit positions. The value of each component in $\mathbf{F}_{\textrm{pro}}$ falls into the range $[0,1]$. Then $\mathbf{F}_{\textrm{pro}}$ is compared to the vector $\mathbf{P}_{\textrm{U}}$, which is randomly generated following the uniform distribution $\mathbf{P}_{\textrm{U}}\sim U(0,1)$. Finally, the BNN outputs the vector $\mathbf{F}_{\textrm{hard}}$, which contains all frozen/information bit positions and it is determined by the comparison results of $\mathbf{F}_{\textrm{pro}}$ and $\mathbf{P}_{\textrm{U}}$, which follows the rule:
\begin{equation}\label{equ.bnn}
  F_{{\textrm{hard}},i}=\left\{
             \begin{array}{lr}
             +1, {F_{{\textrm{pro}},i}} > {P_{\textrm{U},i}} ,  \\
             0,  {F_{{\textrm{pro}},i}} \leq  {P_{\textrm{U},i}}, \\
             \end{array}     
             \forall i=  1, \dots, N
  \right.
\end{equation}
where $+1$ indicates that the $i$-th sub-channel represents an information bit position, while $0$ indicates that the $i$-th sub-channel represents a frozen bit position. In the backward propagation, BNN behaves as an identity function.

Assume that $K'$ out of $N$ sub-channels represent the information bit positions, according to the vector $\mathbf{F}_{\textrm{hard}}$. Then $K'$ information bits of the original bit sequence $\mathbf{b}$ are assigned to these positions, while the remaining $[N-K']$ frozen bit positions are set to zeros, forming a sequence $\mathbf{u}$ of length $N$. {Note that there is a difference between $K'$ and $K$ here, it is explained later in Section \ref{sec.rateadjust}.}
Then, the vector $\mathbf{u}$ is multiplied by the generator matrix $\mathbf{G}_{N}$ to obtain the final encoded bit vector $\mathbf{c}$. The detailed operations are the same as in the conventional algorithm of \cite{8962344Bioglio}. 

{As a benefit of the BNN-aided polar encoder, the learning-based frozen set selection strategy is realized by optimizing the value of each element in the trainable vector $\mathbf{F}_{\textrm{soft}}$, which is driven by the goal of minimizing the loss function, where the WET requirements are considered.} The strategy allows the encoder to adjust the information position probability of each sub-channel according to the practical channel conditions, which guarantees near-optimal WIT performance. Moreover, since the frozen set is selected under both the communication and energy targets, better WIT performance means more resources can be allocated to the energy transfer, then the WET performance gets improved.  


\begin{figure}
	\centering
	\subfloat[]{
	\label{fig.transmitter}
		\includegraphics[width=1.4in]{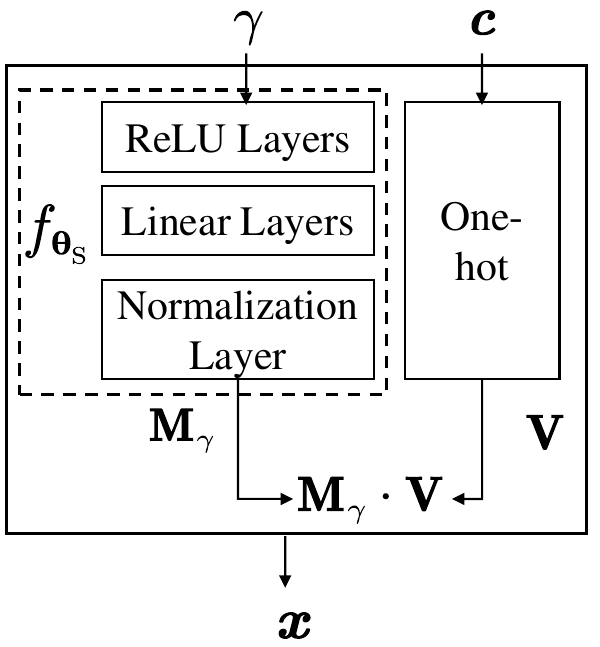}}
	\subfloat[]{
	\label{fig.receiver}
	\includegraphics[width=1.7in]{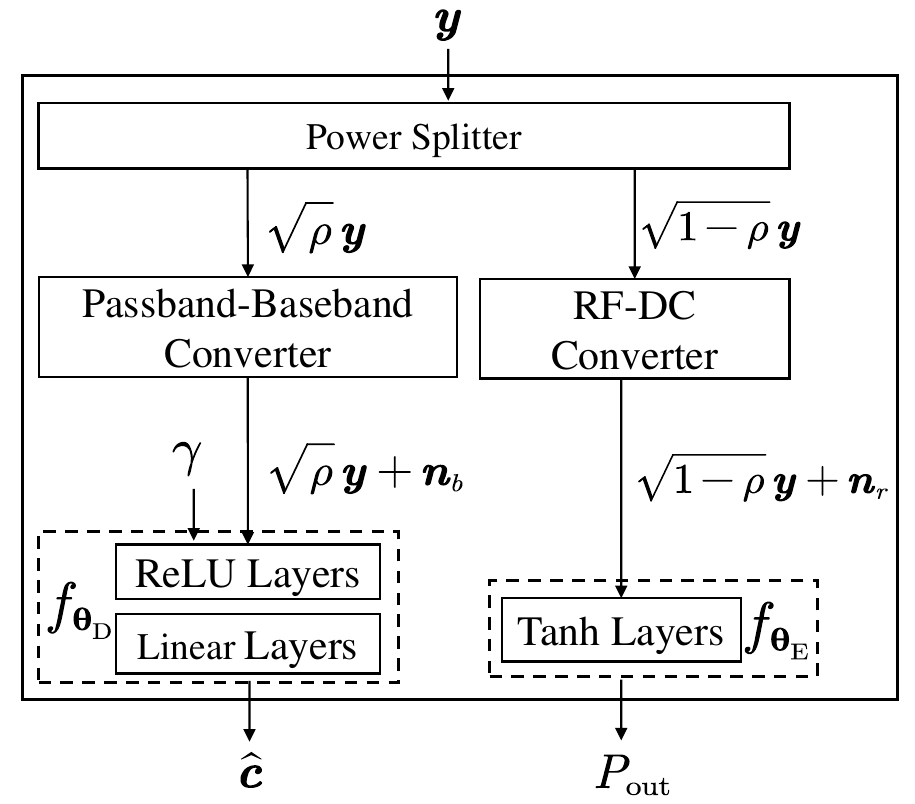}}
	\caption{A generic architecture of (a) AE-Mapper and (b) AE-Demapper and EH.}
	\label{fig.structure}
	\vspace{-0.4 cm}
\end{figure}

\subsubsection{AE-Mapper}
{Inspired by \cite{cammerer2020trainable}, we employed an AE-mapper for our system.}
The AE-mapper of Fig.\ref{fig.systemmodel} is a DNN-based module that designed for modulation. It takes the encoded bits $\mathbf{c}$ as well as the SNR $\gamma$ as its inputs and outputs the modulated vector $\mathbf{x}$. As shown in Fig.3\subref{fig.transmitter}, the modulation process within the AE-mapper of Fig.\ref{fig.systemmodel} mainly consists of two processes, namely the constellation generation and one-hot mapping. The constellation generation is performed on a $I_{\textrm{S}}$-layer DNN $f_{\pmb{\theta}_{\textrm{S}}}$ associated with ReLU and Linear activation functions. The normalization layer is employed as the last layer of the DNN for ensuring that the transmit power $P_\textrm{tr}$ is constrained to a preset value. The DNN takes the SNR $\gamma$ as input and maps it into an $M$-ary constellation set $\mathbf{M}_{\gamma} \in \mathbb{C}^{M \times 2}$, where the two columns represent the real and imaginary parts of the constellation points and $M$ represents the modulation order. The function $f_{\pmb{\theta}_{\textrm{S}}}$ is formulated as
\begin{align}
\mathbf{M}_{\gamma}&=f_{\pmb{\theta}_{\textrm{S}}} (\gamma) \nonumber \\
& =f_{\textrm{norm}} \left(\mathbf{W}^{(\textrm{S})}_{I_{\textrm{S}}}\left [  \cdots f_{\textrm{relu}}(\mathbf{W}^{(\textrm{S})}_1 \gamma+\mathbf{b}^{(\textrm{S})}_1)\cdots\right ]+\mathbf{b}^{(\textrm{S})}_{I_{\textrm{S}}} \right), 
\end{align}
where $f_{\textrm{norm}}(\cdot)$ denotes the normalization function and $f_{\textrm{relu}}(\cdot)$ denotes the ReLU activation function. Furthermore, $\mathbf{W}^{(\textrm{S})}_i$ and $\mathbf{b}^{(\textrm{S})}_i$ jointly constitute the parameter set $\pmb{\theta}_{\textrm{S}}$, which denotes the weights and bias of $i$-th layer of $f_{\pmb{\theta}_{\textrm{S}}}$ for $\forall i= 1, \dots, I_{\textrm{S}}$, respectively.
One-hot mapping is applied to the encoded bit vector $\mathbf{c}$ of size $N$. All the $\log_2 M$ bits in the vector $\mathbf{c}$ are treated as a group and they are mapped into a one-hot vector of size $M \times 1$. Therefore, $[N/\log_2 M]$ one-hot vectors constitute a one-hot matrix $\mathbf{V}\in \mathbb{C}^{M \times N/\log_2 M}$.

The modulated symbols $\mathbf{x}\in \mathbb{C}^{ N/\log_2 M \times 2}$ are obtained by multiplying the constellation set $\mathbf{M}_{\gamma}$ with the matrix $\mathbf{V}$, which represents that each one-hot vector in $\mathbf{V}$ is mapped to the corresponding constellation point in $\mathbf{M}_{\gamma}$. The mapping process is expressed as
\begin{equation}\label{equ.map}
  \mathbf{x}=\textbf{V}^T\cdot \mathbf{M}_{\gamma},
\end{equation}
where $\mathbf{x}$ is the complex baseband symbol set, whose two columns represent the real and imaginary parts of the constellation points, respectively. After obtaining the modulated symbol set $\mathbf{x}$, it is transmitted at the signal power of $P_\textrm{tr}=\mathbb{E}\left[ \|\pmb{x}\|^2 \right]$ either over an AWGN or a Rayleigh channel.

{Overall, the AE-mapper can learn a form of mapping from bit sequences to constellation symbols for given data vs. energy requirements through the training. When the energy requirements change and affect the loss function, the neural network of the AE-mapper will be influenced through back-propagation, thereby affecting the network's output, in other words, the constellation geometry.}
Compared to traditional modulation schemes, whose constellations are fixed, our DNN-based AE-mapper can adjust its output constellations according to the practical information and energy requirements of the receiver and hence improve the WIT and WET performance.

\subsection{Receiver}
The modulated signal $\mathbf{x}$ is transmitted over the complex channel, experiencing both fading and interference, which can be expressed as
\begin{equation}
  \mathbf{y}=\mathbf{h}\odot\mathbf{x}+\mathbf{n}_{0}, 
\end{equation}
where $\mathbf{y}\in \mathbb{C}^{N/\log_2 M \times 1}$ represents the complex-valued received signal and $\odot$ represents element-wise multiplication. Moreover, $\mathbf{h}\in \mathbb{C}^{N/\log_2 M \times 1}$ denotes the channel coefficient, which obeys the complex Gaussian distribution $\mathbf{h}\sim \mathcal{CN}(0,1)$, when it is considered to be a Rayleigh channel. The additive white Gaussian noise $\mathbf{n}_{0}\in \mathbb{C}^{N/\log_2 M \times 1}$ obeys $\mathbf{n}_{0}\sim \mathcal{CN}(0,\sigma^2)$.
Note that the SNR $\gamma$ is defined as $\gamma=\frac{P_\textrm{tr}}{P_\textrm{n}}=\frac{\mathbb{E}\left[ \|\pmb{x}\|^2 \right]}{\mathbb{E}\left[ \|\pmb{n}\|^2 \right]}$, where the noise $\mathbf{n}$ includes the additive white Gaussian noise $\mathbf{n}_{0}$, the antenna noise, the thermal noise and the interference emanating from other users. 

After passing through the channel, the transmitted signal is received by the receiver of Fig.\ref{fig.systemmodel}, which consists of a power splitter, a passband-baseband converter, an RF-DC converter, an EH, an AE-demapper and a hyper-RNN decoder.

\subsubsection{Power Splitter of Fig.\ref{fig.structure}}
The power splitting (PS) method \cite{liu2013wireless} is adopted for information and energy reception. As Fig. \ref{fig.systemmodel} shows, the power splitter takes the received signal $\mathbf{y}$ as its input and then outputs two branches having a specific energy ratio, as determined by the power splitting factor $\rho$. Then the two branches are processed for energy harvesting and information processing, respectively.

As Fig.\ref{fig.systemmodel} shows, before the branches are sent to the AE-mapper and energy harvester, they will first pass through the passband-baseband converter and the RF-DC converter, respectively. The passband-baseband converter converts the passband signals into baseband signals centered at zero frequency, while the RF-DC converter converts the RF signals into DC power. However, extra noise will be introduced during the conversion process. Therefore, the signals at power splitter's output are represented by $\sqrt{\rho}\mathbf{y}+\mathbf{n}_{b}$ and by $\sqrt{1-\rho}\mathbf{y}+\mathbf{n}_{r}$ in Fig.\ref{fig.structure}. Note that for the energy branch, the extra noise $\mathbf{n}_{r}$ is tiny compared to the harvested energy, thus it can be ignored during the energy harvesting process.

\subsubsection{AE-Demapper of Fig.\ref{fig.structure}}\label{sec.receiver}
{The signals $\sqrt{\rho}\mathbf{y}+\mathbf{n}_{b}$ output by the passband-baseband converter are forwarded to the ReLU layer of Fig.\ref{fig.structure} within the AE-demapper, following a similar approach to that of \cite{cammerer2020trainable}.} The AE-demapper $f_{\pmb{\theta}_{\textrm{D}}}$ carries out demodulation by mapping the input signal into the demodulated sequence $\hat{\mathbf{c}}$, which is the prediction of the coded sequence $\mathbf{c}$. In addition to the signal $\sqrt{\rho}\mathbf{y}+\mathbf{n}_{b}$, the SNR $\gamma$ is also fed into the ReLU layer of the AE-demapper as additional environmental information, which can be exploited for improving the demodulation performance.
As depicted in Fig.3\subref{fig.receiver}, a $I_{\textrm{D}}$ layer-DNN that consists of ReLU layers and a Linear layer, is employed for modelling the AE-demapper. The input-output relationship of $f_{\pmb{\theta}_{\textrm{D}}}$ is formulated as follows
\begin{align}
\small
\hat{\mathbf{c}}= & f_{\pmb{\theta}_{\textrm{D}}}([\sqrt{\rho}\mathbf{y}+\mathbf{n}_{b},\gamma])\nonumber \\
=& \mathbf{W}^{(\textrm{D})}_{I_{\textrm{D}}}\left [\cdots f_{\textrm{relu}}\left (\mathbf{W}^{ (\textrm{D})}_1  [\sqrt{\rho}\mathbf{y}+\mathbf{n}_{b},\gamma]+\mathbf{b}^{(\textrm{D})}_1  \right ) \cdots \right ]
+\mathbf{b}^{(\textrm{D})}_{I_{\textrm{D}}}  ,
\end{align}
where again, the $\mathbf{W}^{(\textrm{D})}_i$ and $\mathbf{b}^{(\textrm{D})}_i$ denote the weights and bias of $i$-th layer of $f_{\pmb{\theta}_{\textrm{D}}}$ for $\forall i= 1, \dots, I_{\textrm{D}}$, respectively. {Note that $\hat{\mathbf{c}}$ is a sequence of LLRs instead of binary bits \cite{cammerer2020trainable}, thus it can be smoothly interfaced with the polar decoder.}

{Considering the dependence of the AE-demapper on the SNR, the receiver can obtain the SNR required by the AE-demapper according to \cite{4178493}. However, there may be situations where the receiver has a mismatched SNR. Provided that the difference between the estimated and the actual SNR is small, the system may only suffer from a modest demodulation performance erosion. However, if the difference is significant, there will be a more obvious degradation.}

\subsubsection{EH of Fig.\ref{fig.structure}}\label{sec.EH}
The other signal branch of $\sqrt{1-\rho}\mathbf{y}+\mathbf{n}_{r}$ having the RF power of $P_{\textrm{in}}$ is processed by the Tanh layer of Fig.3\subref{fig.receiver} for energy harvesting. The non-linear EH model of \cite{varasteh2020learning} is adopted. As Fig.3\subref{fig.receiver} shows, it is modeled by a $I_{\textrm{E}}$-layer DNN relying on the Tanh activation function, which is represented by $f_{\pmb{\theta}_{\textrm{E}}}$ associated with the parameter set $\pmb{\theta}_{\textrm{E}}$. The DNN-aided EH model adopts the classic non-linear regression algorithm to harvest energy, which conforms to the non-linear characteristic of practical EHs. 
The function $f_{\pmb{\theta}_{\textrm{E}}}$ is expressed as \cite{varasteh2020learning}:
\begin{align}\label{equ.EH}
  P_{\textrm{out}}=&f_{\pmb{\theta}_\textrm{E}}(P_{\textrm{in}}) \nonumber \\
  =&f_{\textrm{tanh}}(\mathbf{W}^{(\textrm{E})}_{I_{(\textrm{E})}}\left [\cdots f_{\textrm{tanh}}(\mathbf{W}^{(\textrm{E})}_1\cdot P_{\textrm{in}}+\mathbf{b}^{(\textrm{E})}_1)\cdots \right ]
+\mathbf{b}^{(\textrm{E})}_{I_{(\textrm{E})}}),
\end{align}
where $f_{\textrm{tanh}}(\cdot)$ represents the Tanh activation function and $P_{\textrm{in}}=(1-\rho)\|\pmb{y}\|^2$, since the power of noise $\mathbf{n}_{r}$ is ignored. In line with the information signal model, in the energy model, $\mathbf{W}^{(\textrm{E})}_i$ and  $\textbf{b}^{(\textrm{E})}_i$ denote the weights and bias of $i$-th layer of $f_{\pmb{\theta}_{\textrm{E}}}$ for $\forall i= 1, \dots, I_{\textrm{E}}$, respectively. Moreover, the non-linear characteristic of EH is learnt by training relying on a large amount of data collected in advance and {we directly adopt the pre-trained model provided by \cite{varasteh2020learning} in our work.} Therefore, the EH relies on a simple mapping function having fixed parameters during the end-to-end global optimization process, without adjusting itself. 

\begin{figure}
  \centering
  \includegraphics[width=3.1in]{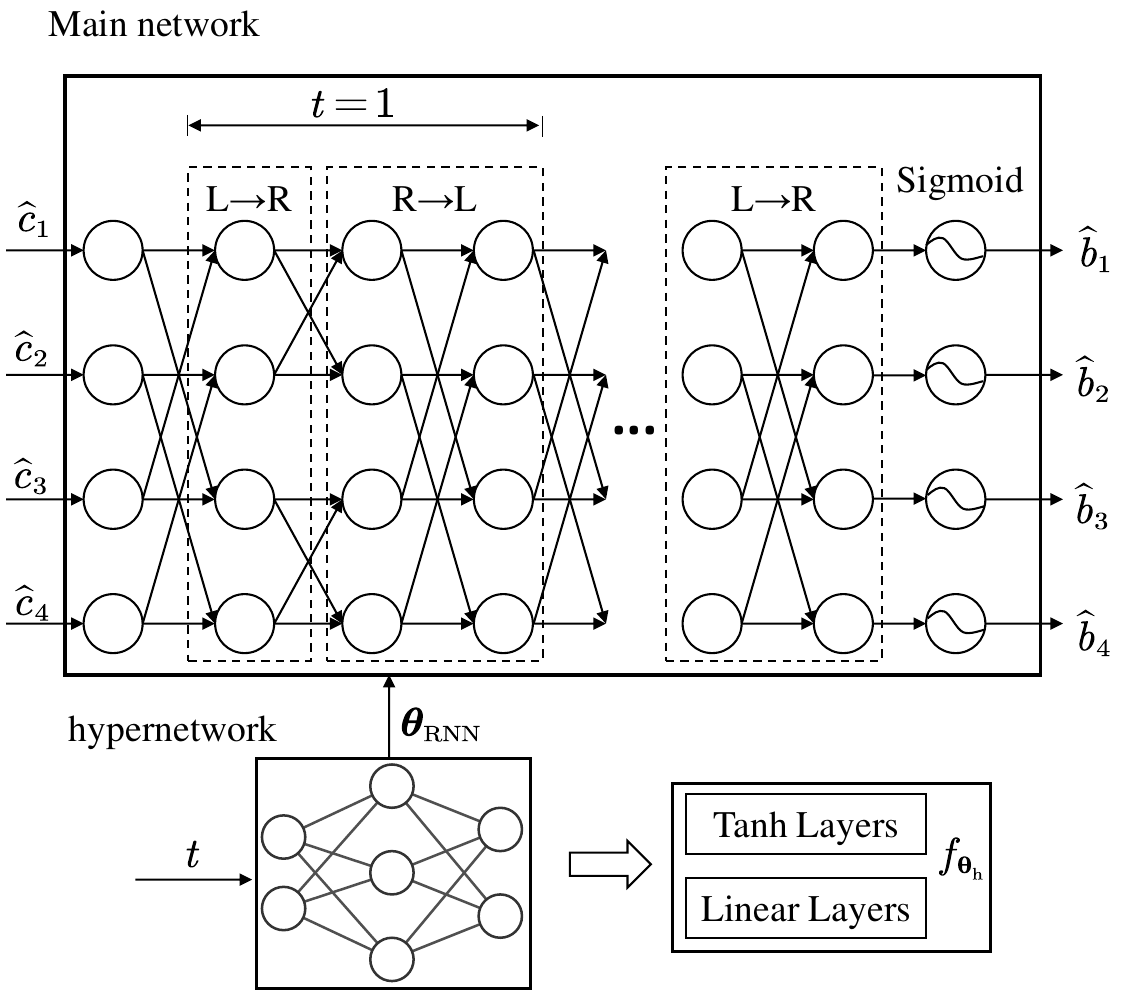}\\
  \caption{An example of hyper-RNN decoding with $N=4$.}\label{fig.BP}
\end{figure}

\subsubsection{Hyper-RNN Decoder of Fig.\ref{fig.BP}}\label{sec.decoder}

\begin{figure}
  \centering
  \includegraphics[width=3in]{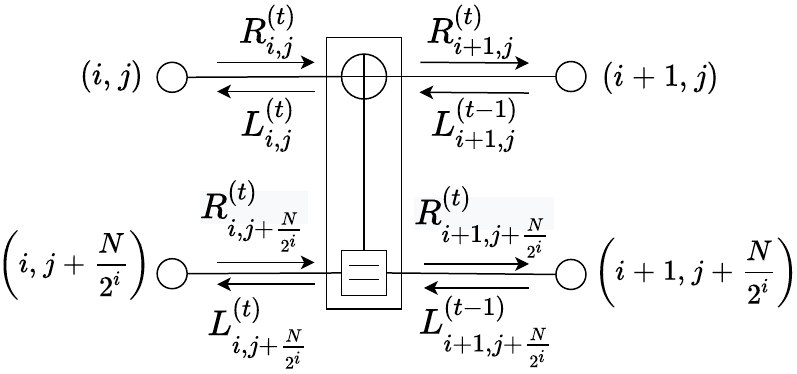}\\
  \caption{Processing element of the polar code graph.}\label{fig.PE}
\end{figure}
The hyper-RNN decoder performs channel decoding. {Considering that the parallel structure of BP decoding is compatible with neural networks, and the BNN-aided encoder design is tailored to iterative decoding \cite{ebada2019deep}, thus we harnessed BP decoding.} 
In contrast to the pure DNN-based BP decoder of \cite{8109997}, our hyper-RNN decoder is based on the structure of hypernetworks, which includes a pair of networks, namely a main network based on the RNN and a hypernetwork. {In contrast to the DNN where neurons have multiple weights for different number of iterations, the RNN’s network weights for different number of iterations are shareable, which reduces the memory requirement and it is an advantage of the RNN decoder \cite{teng2019low }.}

As shown in Fig.\ref{fig.systemmodel},
the main network carries out the BP decoding process, which is based on the Forney-style factor graph of \cite{arikan2010graph}. 
{Assume that the number of decoding iteration is $T$,} the main network processes the demodulated vector $\hat{\mathbf{c}}$ as its input and outputs the prediction $\hat{\mathbf{b}}$ of the original bits $\mathbf{b}$, while the hypernetwork takes the number of iterations $t$ as its input and outputs the corresponding weights $\pmb{\theta}_{\textrm{RNN}}$ of each layer for the main network. {It is worth noting that $T$ is a constant while $t$ is a variable and we have $t= 1,2, \dots, T$ during the decoding process.}
Thus it helps the decoder adapt to diverse decoding situations associated with different number of iterations and achieve optimal decoding performance. 
By contrast, the commonly used pure DNN-based decoder does not take into account the number of iterations as its input to generate the corresponding network parameters, thus it cannot achieve optimal performance for different numbers of decoding iterations. Having said that, it can still attain satisfactory performance. This comparison demonstrates the superior adaptability of our hyper-RNN decoder over the pure DNN-based decoder.

Further details of the hyper-RNN decoder are seen in Fig. \ref{fig.BP}, where the hypernetwork $f_{\pmb{\theta}_{\textrm{h}}}$ is represented by a $I_{\textrm{h}}$-layer DNN associated with Tanh and Linear activation functions. It takes the number of iterations $t$ as its input and generates the corresponding weights for the main network. The function $f_{\pmb{\theta}_{\textrm{h}}}$ is formulated as 
\begin{align}
\pmb{\theta}_{\textrm{RNN}}= & f_{\pmb{\theta}_{\textrm{h}}}(t)\nonumber \\
=& \mathbf{W}^{(\textrm{h})}_{I_{\textrm{h}}}\left [\cdots f_{\textrm{tanh}}\left (\mathbf{W}^{ (\textrm{h})}_1  t  \right ) \cdots \right ]  ,
\end{align}
where $\mathbf{W}^{(\textrm{h})}_i$ denotes the weights of $i$-th layer of $f_{\pmb{\theta}_{\textrm{h}}}$ for $\forall i= 1, \dots, I_{\textrm{h}}$.

As portrayed in Fig. \ref{fig.BP}, the main network $f_{\pmb{\theta}_{\textrm{RNN}}}$ that carries out the decoding process is based on a multi-layer partially-connected RNN, which has a similar structure to that of the polar code's factor graph. Consider the polar code of length $N$ as an example. The RNN contains $\log_2 N$ stages and $[\log_2 N + 1]$ layers, where each stage includes $N/2$ basic computation units termed as processing-elements (PEs), which are shown in Fig.\ref{fig.PE}. As the number of decoding iterations $T$ increases, the neural network expands correspondingly by repeating the initial $[\log_2 N + 1]$ layers $T$ times, with $[2(\log_2 N-1)T+1]$ hidden layers in total.
The Sigmoid function is applied in the last layer of the RNN to ensure that the output falls into the range of $[0,1]$.
The RNN takes the demodulated vector $\hat{\mathbf{c}}$ as its input and outputs the prediction $\hat{\mathbf{b}}$. The function of the RNN $f_{\pmb{\theta}_{\textrm{RNN}}}$ is formulated as
\begin{align}
    \hat{\mathbf{b}}=f_{\pmb{\theta}_{\textrm{RNN}}}( \hat{\mathbf{c}}).
\end{align}
More specifically, the decoding process is based on the transmission and on the update of logarithmic likelihood ratios (LLRs) of each neuron in the RNN. Two types of LLRs, namely the left-to-right (L → R) LLR $R_{i,j}^{(t)}$ and the right-to-left (R → L) LLR $L_{i,j}^{(t)}$, are propagated along the connections between neurons. The initial values of these two LLRs are defined as
\begin{equation}\label{equ.LLRs}
  R_{1,j}^{(1)}=\left\{
             \begin{array}{lr}
             0, \ \ \ \ \ j\in \mathbf{F}_{\textrm{hard}} \\
             +\infty, \ \ j\notin \mathbf{F}_{\textrm{hard}}
             \end{array}
  \right.
  \hspace{1.5em}
  L_{n+1,j}^{(1)}=\hat{\mathbf{c}},
\end{equation}
where $R_{i,j}^{(t)}$ and $L_{i,j}^{(t)}$ denote the left-to-right LLR and the  right-to-left LLR of the $j$-th neuron in the $i$-th stage at the $t$-th iteration, respectively. The input $ \hat{\mathbf{c}}$ is an LLR vector and it denotes the initial right-to-left LLR. In the RNN, the updates of $R_{i,j}^{(t)}$ and $L_{i,j}^{(t)}$ depend upon their positions. Therefore, the function of the RNN $f_{\pmb{\theta}_{\textrm{RNN}}}$ also represents the LLR updating process, which can be formulated as follows:
\begin{flalign} 
  &\ f_{\pmb{\theta}_{\textrm{RNN}}}= \nonumber &
\end{flalign}
\begin{align}\label{equ.scaledbp} 
  \left\{
             \begin{array}{lr}
             L_{i,j}^{(t)}=\alpha_{i,j}\cdot g(L_{i+1,j}^{(t-1)},L_{i+1,j+N/{2^i}}^{(t-1)}+R_{i,j+N/{2^i}}^{(t)}),  \\
             L_{i,j+N/{2^i}}^{(t)}=\alpha_{i,j+N/{2^i}}\cdot g(R_{i,j}^{(t)},L_{i+1,j}^{(t-1)})+L_{i+1,j+N/{2^i}}^{(t-1)}, \\
             R_{i+1,j}^{(t)}=\beta_{i+1,j}\cdot g(R_{i,j}^{(t)},L_{i+1,j+N/{2^i}}^{(t-1)}+R_{i,j+N/{2^i}}^{(t)}), \\
             R_{i+1,j+N/{2^i}}^{(t)}=\beta_{i+1,j+N/{2^i}}\cdot g(R_{i,j}^{(t)},L_{i+1,j}^{(t-1)})+R_{i,j+N/{2^i}}^{(t)},
             \end{array}
  \right.
\end{align}

where we have $g(a,b)\approx \textrm{sign}(a)\textrm{sign}(b)\min(|a|,|b|)$, while $\alpha_{i,j}$ as well as $\beta_{i,j}$  denote the right-to-left and the left-to-right scaling parameters of the $j$-th neuron at the $i$-th stage, respectively. Moreover, they are also considered as the weights of the RNN and are incorporated into the parameter set $\pmb{\theta}_{\textrm{RNN}}$. Since the weights of each layer are shareable among different iterations in RNNs, $\alpha_{i,j}$ and $\beta_{i,j}$ are independent of the number of iterations $t$, which substantially reduces the parameter storage requirements. By contrast, the weights of each layer are different for different iterations in conventional DNNs. This is an inherent impediment of the pure DNN-based decoder, compared to the RNN-based decoder.

\section{End-to-End Design}\label{sec.train}
In this section, the optimization objective is formulated by considering the WIT performance, WET performance as well as the code rate adjustment. Then we provide an example to introduce the end-to-end training process and the loss function design in detail.

\subsection{Optimization Problem}\label{subsec.op}
We aim for satisfying the energy harvesting requirements $P_{\textrm{targ}}$, while minimizing the BER without any transmission rate erosion. Driven by this, we formulate the optimization objective, which includes the WIT part, the WET part and the rate adjustment part.
\subsubsection{WIT part}
The WIT part mainly aims for guaranteeing the communication performance of the IDEN system. The binary cross entropy (CE) is harnessed for characterizing the difference between the original bits $b_i$ and their prediction $\hat{b}_i$, for $\forall i= 1, \dots, N$:
\begin{align}
\mathcal{L}_{\textrm{WIT}}=\sum_{i=1}^N{\left( b_i\log \hat{b}_i+\left( 1-b_i \right) \log \left( 1-\hat{b}_i \right) \right)}.
\end{align}

\subsubsection{WET part}
The WET part represents the impact of energy transfer on the network's optimization. In this paper, a dynamic power splitting scheme is designed, which allows the power splitter to adjust the power splitting factor $\rho$ according to the practical information and energy requirements. Since it is in an IDEN system, there is a trade-off between the WIT and WET, which should be considered as part of our optimization problem. Moreover, a target of the harvested energy $P_{\textrm{targ}}$ is given beforehand and it is expected that the harvested energy $P_{\textrm{out}}$ converges to the given energy target during the training process.
Therefore, the difference between the harvested energy and the target, as well as the trade-off between the data and energy are both taken into consideration when designing the optimization objective. Consequently, the WET part is formulated as follows:
\begin{align}\label{eq.WETpart}
\mathcal{L}_{\textrm{WET}}=\underbrace{\frac{\lambda}{P_{\textrm{out}}}}_{\textrm{Part 1}}+\underbrace{\left(P_{\textrm{targ}}-P_{\textrm{out}}\right)+\left(P_{\textrm{targ}}-P_{\textrm{out}}\right)^2}_{\textrm{Part 2}},
\end{align}
where $\lambda$ is a non-negative bias parameter and it is introduced to depict the trade-off between the information requirements and the energy harvesting requirements.
As Eq. \eqref{eq.WETpart} shows, the WET part includes two subparts.
Part 1 is jointly determined by both the bias parameter $\lambda$ and the harvested energy $P_{\textrm{out}}$. A larger $\lambda$ indicates that the IDEN has to harvest more energy at the cost of degraded BER performance. Part 2 is formulated by the the first and second moments of the difference between the harvested energy $P_{\textrm{out}}$ and the target energy $P_{\textrm{targ}}$, which ensures that the harvested energy is no less than the target energy, {while converging to the target energy and achieving it approximately.} 
Thereby, the system can achieve the optimal WET performance under a certain trade-off. 
{In order to increase $P_{\textrm{out}}$, note that during training, we set a subtraction term $(P_{\textrm{targ}}-P_{\textrm{out}})$. Although we also set a squared term $(P_{\textrm{targ}}-P_{\textrm{out}})^2$ to prevent $P_{\textrm{out}}$ from becoming too large and exceeding the target, it is still possible that $P_{\textrm{out}}$ exceeds the target and $\mathcal{L}_{\textrm{WET}}$ becomes negative during training. However, regardless whether $\mathcal{L}_{\textrm{WET}}$ is positive or negative, it is able to push the output $P_{\textrm{out}}$ close to the target energy, so as to achieve the learning objective. }

\subsubsection{Rate adjustment}\label{sec.rateadjust}
The rate adjustment part is designed for constraining the code rate to the target code rate $R_\textrm{targ}$. According to the characteristics of the BNN described in Section \ref{sec.polar encoder}, the location and number of frozen/information bits are dynamically adjusted during the training process. If the trend of the code rate is not constrained and optimized, the optimization objective of minimizing the communication BER will force the rate to converge to a very low code rate, which may result in a waste of resources. Therefore, it is essential to take the code rate optimization into consideration. During the training process, the average code rate $R$ is calculated by
\begin{equation}
  R=\frac{1}{N}\sum_{i=1}^N{f_{\textrm{sig}}(F_{\textrm{soft},i})},
\end{equation} 
The target code rate is set to $R_\textrm{targ}=K/N$, where $K$ denotes the number of information bits. It is expected that the average code rate $R$ converges to the preset target rate $R_\textrm{targ}$ through training. Thus, the optimization objective is  formulated by the second moments of the difference between $R$ and  $R_\textrm{targ}$ as follows
\begin{align}\label{eq.WET part}
\mathcal{L}_{\textrm{rate}}=&\left(R-R_\textrm{targ}\right)^2  \nonumber\\
=&\left(\frac{1}{N}\sum_{i=1}^N{f_{\textrm{sig}}(F_{\textrm{soft},i})}-\frac{K}{N}\right)^2,
\end{align}
The rate adjustment part guarantees the IDEN performance, while avoiding the loss of transmission rate.

In general, the number of information bits $K'$ during training is determined by Eq.\eqref{equ.bnn} of Section \ref{sec.polar encoder}. It is dynamically varying and hence it is not necessarily equal to the intended number $K$. However, as a benefit of the optimization of the code rate (which is also an indirect optimization of $\mathbf{F}_{\textrm{soft}}$), $K'$ approaches the target $K$, thus the target can be approached. In the subsequent test process, in order to reach the target $R_\textrm{targ}$, the information bit positions are selected as the positions corresponding to the largest $K$ elements in $\mathbf{F}_{\textrm{soft}}$, while the remaining positions are the frozen bit positions.

Based on the above, the objective function (OF) of the optimization problem can be formulated as follows
\begin{align}\label{pf-1}
    \textrm{(P1):}\underset{\pmb{\theta}_\textrm{S},\pmb{\theta}_\textrm{D},\pmb{\theta}_{\textrm{RNN}},\pmb{\theta}_\textrm{h},\rho}{\text{min}}&  \textrm{E} \left [\mathcal{L}_{\textrm{WIT}}+\mathcal{L}_{\textrm{WET}}+\mathcal{L}_{\textrm{rate}}  \right ] 
    \\
    \tag{\ref{pf-1}{a}}\label{eq.Pdlconstraint}
    \text{s.t.}~ & (2), (5), (6), (7) ~\textrm{and}~ (8) , \\
    \tag{\ref{pf-1}{b}}\label{eq.Ptrconstraint}
    &\mathbb{E}\left[ \|\pmb{x}\|^2 \right]\leq P_{\textrm{tr}}, \\
    \tag{\ref{pf-1}{c}}\label{eq.Psconstraint}
    & 0\leq \rho \leq 1,
\end{align}
where $P_{\textrm{tr}}$ represents the transmit power constraint and $\rho$ denotes the power splitter factor. 
Eq.\eqref{eq.Pdlconstraint} represents that the end-to-end IDEN process follows the mapping relationship of each neural network module. Eq.\eqref{eq.Ptrconstraint} formulates the transmit power constraint, while Eq.\eqref{eq.Psconstraint} represents the constraint on the power splitting factor.

\subsection{End-to-End Training}
Before using the proposed end-to-end polar 
coded IDEN system, we have to train it offline for desirable performance. Fig.\ref{fig.systemmodel} demonstrates the architecture of the end-to-end IDEN system, where the original bit sequence $\mathbf{b}$ is generated randomly and sent to the BNN-aided polar encoder. Once the encoder generates the frozen bit positions and constructs the coded bit vector $\mathbf{c}$, the vector $\mathbf{c}$ together with the SNR $\gamma$ are forwarded to the AE-mapper of Fig.\ref{fig.systemmodel}. The AE-mapper designs the complex modulation symbols to satisfy both the communication and energy requirements. Then the modulated signal $\mathbf{x}$ is transmitted through either an AWGN or a Rayleigh channel. The received signal $\mathbf{y}$ is first divided into two branches by the power splitter having the power splitting factor $\rho$. After passing through the RF-DC converter and the passband-baseband converter respectively, the branch $\sqrt{1-\rho}\mathbf{y}+\mathbf{n}_{r}$ of Fig.\ref{fig.structure} is sent to the EH for energy harvesting, while the other branch $\sqrt{\rho}\mathbf{y}+\mathbf{n}_{b}$ is forwarded to the AE-demapper to obtain the demodulated signal $\hat{\textbf{c}}$. Then the signal $\hat{\textbf{c}}$ is forwarded to the RNN of the hyper-RNN decoder, while the number of iterations $t$ is fed into the hypernetwork to assist in the training of the decoder. Assume that the number of training iterations is set to $T_{\textrm{train}}=T$. Then an input set of $\mathbf{t}=[1,2,\cdots,T]$ will be sent to the hypernetwork for training to ensure that the hypernetwork can generate the weights of the RNN that are optimal for each specific iteration $t_i$, for $\forall i=1,2,\cdots,T$. As a benefit of training, the hyper-RNN decoder can be adaptive to all decoding situations where $T_{\textrm{test}} \leq T$, and achieve satisfactory decoding performance.
Finally, the decoder outputs a bit vector $\hat{\mathbf{b}}$ as the prediction of the original bit sequence $\mathbf{b}$. 

The training procedure relies on minimizing the difference between the prediction $\hat{\mathbf{b}}$ and the original bits $\mathbf{b}$, while meeting both the energy harvesting target and the code rate target. Therefore, the loss function is formulated as follows:
\begin{align}\label{equ.loss}
  \mathcal{L}(\textbf{b},\hat{\textbf{b}},\pmb{\theta},\mathbf{F}_{\textrm{soft}})=&-\frac{1}{|\mathcal{B}|}\sum_{\textbf{b}\in |\mathcal{B}|}\Big(\mathcal{L}_{\textrm{WIT}}    +\beta_1\frac{\lambda}{P_{\textrm{out}}(\textbf{b})}  \nonumber \\
  &+\beta_2\left(P_{\textrm{targ}}-P_{\textrm{out}}\right)+\beta_3\left(P_{\textrm{targ}}-P_{\textrm{out}}\right)^2  \nonumber \\
  &+\beta_4\mathcal{L}_{\textrm{rate}}
   \Big),
\end{align}

where $|\mathcal{B}|$ denotes the batchsize of the data harnessed for training and $\beta_i$ for $\forall i= 1, \dots, 4$ denote the weighting factors used for adjusting the impact of each component on the objective. {Since the values of $\beta_i$ are not fixed and they are easily affected by the data vs. energy requirements, they are obtained through extensive Monte-Carlo simulations in our simulations.}
As Eq. \eqref{equ.loss} shows, the loss function is the weighted sum of multiple components that contains the WIT part, WET part and rate adjustment part. In summary, the loss function used for training corresponds to the OF of (P1). 
Moreover, in order to obtain the optimal network parameters, {the end-to-end joint optimization solution depends on the channel statistics,} and we harness the stochastic gradient descent algorithm (SGD) \cite{goodfellow2016deep} for updating the parameters during the training process.

{In addition, we analyze the complexity of the joint optimization algorithm and express it by the Big-O complexity. We assume that each hidden layer of the AE-mapper, the AE-demapper, the EH and the hypernetwork contains $k_\textrm{S}$, $k_\textrm{D}$, $k_\textrm{E}$ and $k_\textrm{h}$ neurons, respectively. Therefore, the Big-O complexity is $\mathcal{O}(k_\textrm{S}^2I_\textrm{S}+k_\textrm{S}M+k_\textrm{D}^2I_\textrm{D}+k_\textrm{D}log_2M+k_\textrm{h}TN\textrm{log}N+k_\textrm{h}^2TI_\textrm{h}+k_\textrm{E}^2I_\textrm{E})$. We can observe that many parameters (e.g., the number of network layers, number of neurons, code length, etc.) affect the overall complexity, thus the coordination between the modules is important in the end-to-end design.}

\section{Simulation Results}

\subsection{Simulation Setup}
We characterize the proposed deep learning aided IDEN system by comparing it to the conventional end-to-end IDEN system in terms of the local as well as global performance of both the communication and energy transfer. The local performance includes the BNN-aided encoder's performance in terms of WIT and WET, as well as the hyper-RNN decoder's performance in terms of WIT, while the global performance represents the WIT and WET performance of the whole IDEN system. The simulations are carried out based on the platform Tensorflow 1.14. 
In our simulations, {since all modules in the proposed IDEN system are based on DNNs, and the complexity and latency of the decoding module are affected both by the number of iterations $T$ and by the code length $N$, long codes may lead to insufficient computer memory and affect the joint training of the system. Therefore, we rely on polar codes associated with $N=64$ to facilitate our  system performance verification.} 
{$R_\textrm{targ}$ is set as $1/2$. To ensure that the value of $P_\textrm{targ}$ is appropriate, the value range is set to satisfy the linear region as well as the saturated region of the EH according to the nonlinear characteristic of EH model in \cite{varasteh2020learning}.}
The modulation order is set to $M = 4$, and the Adam optimizer is implemented with the learning rate set to $\delta=0.005$. In addition, {we choose the AWGN channel and the Rayleigh channel for signal transmission because they are classic best and worst-case channel types, which are convenient for verifying the feasibility of the end-to-end learning idea for the fully DNN-based IDEN system.}
We also set the noise power $P_\textrm{n}$ for different channels, e.g., 
$P_\textrm{n}=-3$ dBm for AWGN channels and $P_\textrm{n}=-13$ dBm for Rayleigh channels.



\begin{figure}[t]
	\centering  
	\subfloat[AWGN]{
	\label{fig.local1}
		\includegraphics[width=3in]{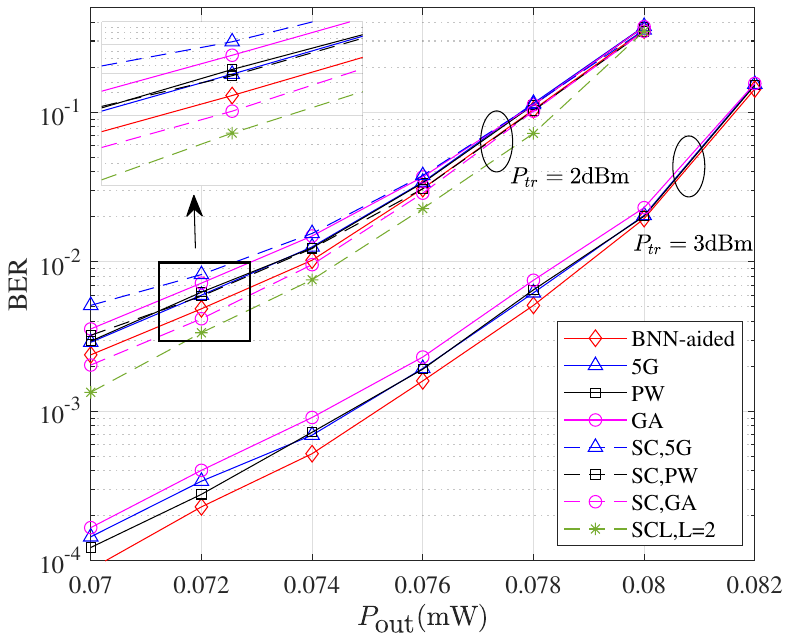}}\\
	\subfloat[Rayleigh]{
	\label{fig.local1_R}
	\includegraphics[width=3in]{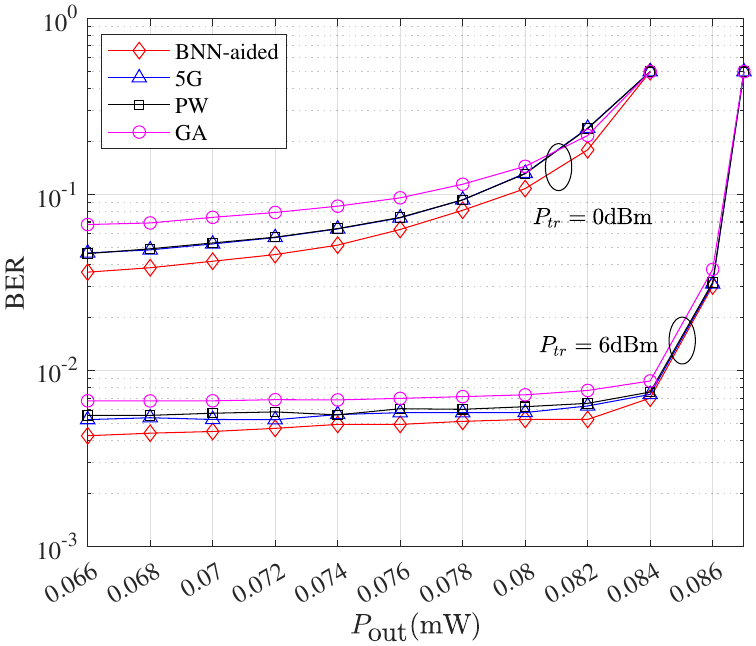}}
	\caption{BER performance comparisons between the BNN-aided polar encoder and the conventional polar encoders with BP, SC, SCL decoders. (a) Performance over AWGN channels using 50 decoding iterations under different transmit power. (b) Performance over Rayleigh channels using 50 decoding iterations under different transmit power.}
	\label{fig.local1AR}
\end{figure}

\subsection{Local Performance Comparison}
In order to show the advantages of our BNN-aided polar encoder and the hyper-RNN decoder in WET and WIT in a more detailed way, we commence by conducting the simulations for local performance comparisons, respectively.

Fig.\ref{fig.local1AR} demonstrates the WIT and WET performance of our BNN-aided polar encoder and of conventional polar encoders for different construction algorithms, like GA, Polarization weight (PW) and the bit-channel reliability sequence proposed by 5G standardization \cite{3rd2021technical}. {We also apply different decoders, such as BP decoders, successive cancellation (SC) decoders and SCL decoders without CRC-aided, to verify the impact of decoder performance.} The simulations are conducted over both the AWGN and Rayleigh channels at different transmit powers $P_\textrm{tr}$. Again, the modulation scheme is $4$-QAM and the BP decoding algorithm is adopted. {The number of BP iterations is set to $50$ for both the AWGN channel and the Rayleigh channel, and the list size of the SCL decoder is set as $2$}. 
Firstly, it can be observed from Fig.6\subref{fig.local1} that the BER curves of all polar encoders are shifted upwards upon reducing the transmit power. Moreover, the BER performance deteriorates as the harvested energy $P_\textrm{out}$ is increased, which demonstrates the inherent trade-off between WIT and WET in IDEN system. {In the group using the BP decoder, i.e. the solid lines in Fig.\ref{fig.local1AR}, our BNN-aided polar encoder exhibits the lowest BER at a given transmit power.} Viewed from another perspective, our BNN-aided polar encoder can harvest more energy  than its counterparts at the same BER. This can be attribute to the BNN-aided polar code construction algorithm, which adjusts the frozen bit positions during the training process according to the variable channel conditions. Thus the information bits can be transmitted under better sub-channel conditions in the IDEN scenarios, hence achieving improved communication performance. Meanwhile, increased communication resources can be allocated to WET services to harvest more energy. 
By contrast, the conventional construction algorithms cannot modify the  polar code construction in real time according to the feedback of information and energy performance under varying energy requirements and channel conditions. However, compared to its SC/SCL based counterparts represented by the dashed lines in Fig.6\subref{fig.local1}, our BNN-aided encoder's performance becomes sub-optimal. We can observe that its SC based counterparts perform similarly to the $50$-iteration BP based counterparts, while the SCL based solution has the most obvious advantage. This is mainly due to the inferior performance of the BP decoder compared to the SC/SCL decoders.
Based on this, we can find that the gain attained by the BNN aided encoding algorithm is limited, and the overall performance of the system is also affected by the type of decoder that cannot be ignored. Therefore, this also reveals that constructing the SC/SCL based BNN-aided encoder is promising.  

As seen from Fig.6\subref{fig.local1_R}, for a Rayleigh channel, the BER curves follow  similar trends as in AWGN channels. {We can observe that the BER curves associated with $P_{\textrm{tr}} = 6$ dBm remain stable when the harvested energy ranges from $0.066$ mW to $0.084$ mW, but rise sharply when the harvested energy is higher (around $0.084$ mW to $0.087$ mW). This is because the transmit power is high, thus a small variation of the power splitting factor used for WET $(1-\rho) \in[0.08,0.2]$ can lead to a noticeable variation of the harvested energy. However, for the WIT, a small change in the power splitting factor results in an unnoticeable change in the BER performance, thus the BER curves remain nearly unaltered. The curves rise sharply in the region $[0.084, 0.087]$mW, mainly due to the non-linear characteristic of the energy harvester. In the saturated region, the output energy increases slowly when the input energy is high. Therefore, in order to achieve the same amount of energy growth as before ($\varDelta  P_\textrm{out}=0.002$ mW), the power splitter has to allocate more resources to the WIT, thus the communication performance deteriorates dramatically.}
Additionally, our BNN-aided polar encoder achieves a lower BER at the same harvested energy target compared to its traditional counterparts with BP decoders, demonstrating its advantage in both WET and WIT for transmission over a Rayleigh channel. 


\begin{figure}[h]
	\centering  
	\subfloat[AWGN]{
	\label{fig.local2}
		\includegraphics[width=3in]{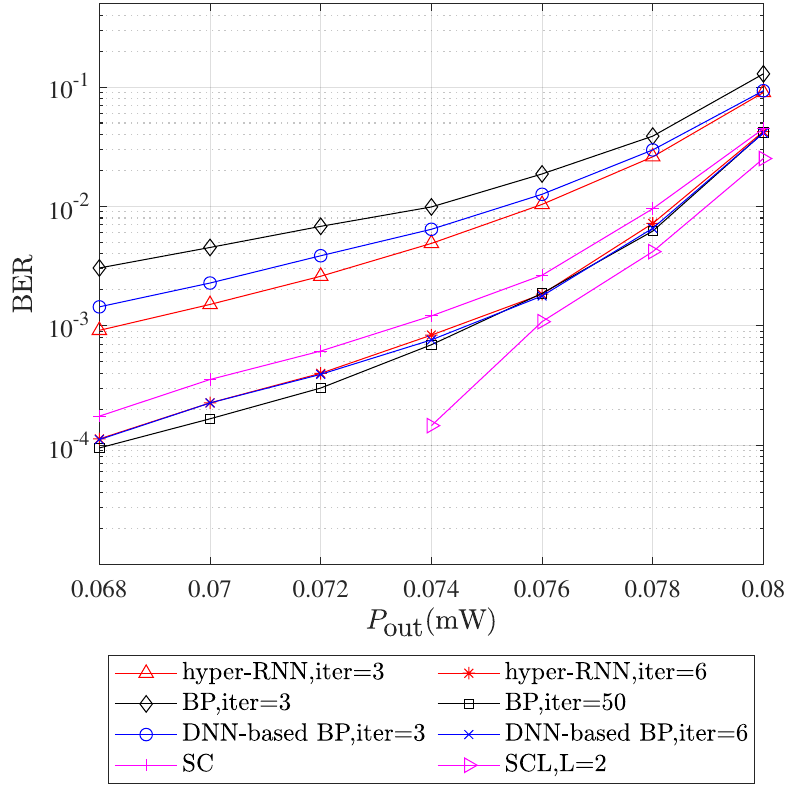}}\\
	\subfloat[Rayleigh]{
	\label{fig.local2_R}
	\includegraphics[width=3in]{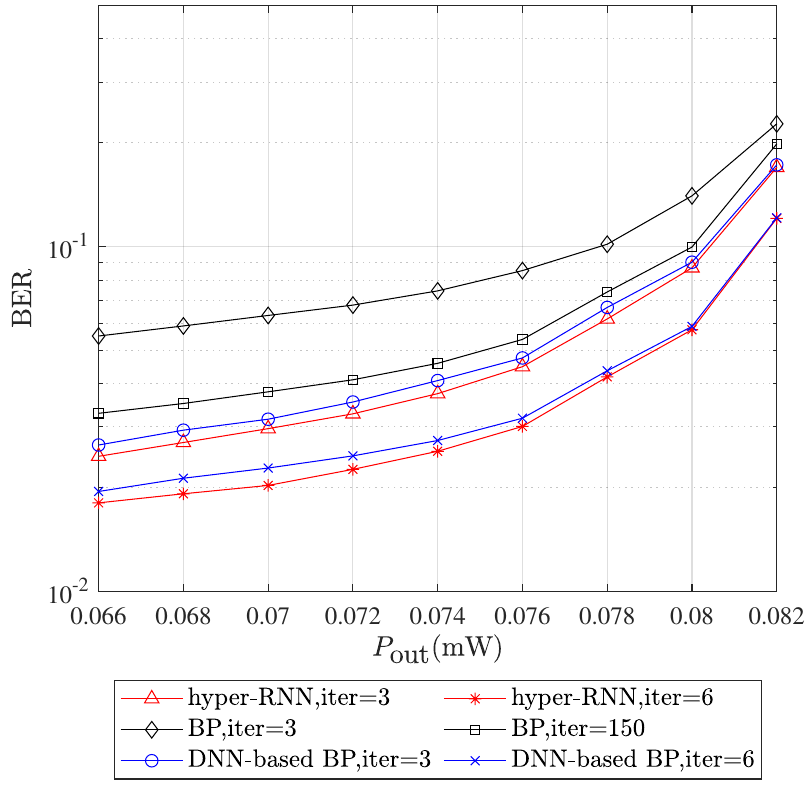}}
	\caption{BER performance comparisons between different decoders. (a) Performance over AWGN channels with $P_\textrm{tr}=3$dBm under different iterations. (b) Performance over Rayleigh channels with $P_\textrm{tr}=0$dBm under different iterations.}
	\label{fig.local2AR}
\end{figure}

\begin{figure}
  \centering
  \includegraphics[width=2.8in]{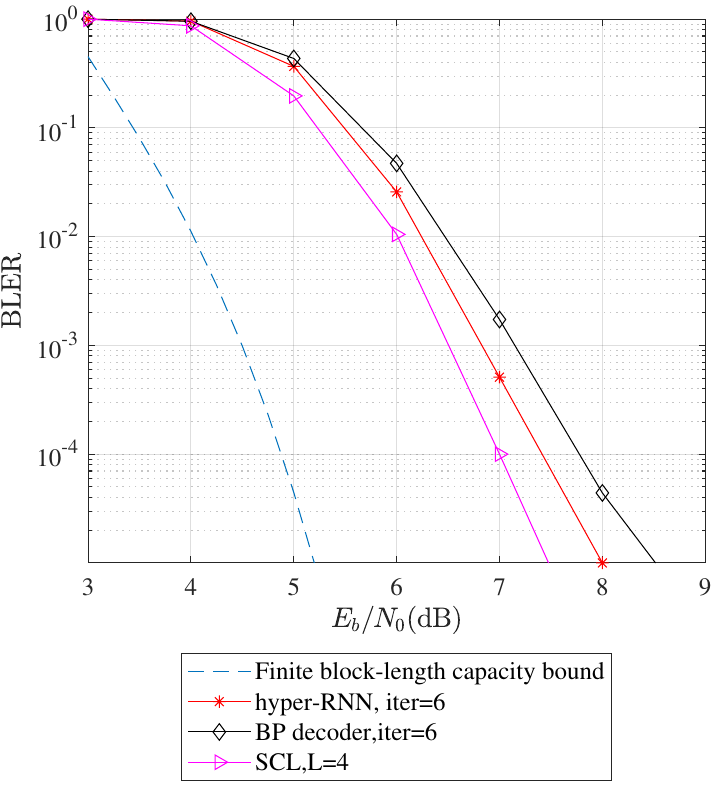}\\
  \caption{BLER performance of the hyper-RNN decoder, the conventional BP decoder and the SCL decoder with $P_\textrm{out}=0.078$mW over an AWGN channel.}\label{fig.finiteblock}
\end{figure}

We also compare the BER performance of the proposed hyper-RNN decoder to that of the conventional BP decoder and the DNN-based decoder. {In addition, we add extra SC and SCL decoders without CRC in AWGN channels for comparison.} The simulations are carried out using $4$-QAM at $P_{\textrm{tr}}=3$ dBm for an AWGN channel and $P_{\textrm{tr}}=0$ dBm for a Rayleigh channel.
As shown in Fig.\ref{fig.local2AR}, our proposed hyper-RNN based decoder achieves lower BER than the conventional BP decoder at the same number of decoding iterations under both the AWGN channels and the Rayleigh channels. In particular, {the 6-iteration hyper-RNN decoder performs similarly to the 50-iteration BP decoder under an AWGN channel, demonstrating its superiority in polar decoding. This is due to the beneficial characteristics of the RNN, whose neurons' weights  $\alpha_{i,j}$, $\beta_{i,j}$ in Eq.\ref{equ.scaledbp} can be used as the scaling parameters, which lend themselves to training. Thus the decoder achieves its optimal BER performance after training. Moreover, the 6-iteration hyper-RNN decoder outperforms the SC decoder, but it is inferior to the SCL decoder associated with $L=2$. This reflects the weakness of BP decoder in decoding.}
Meanwhile, Fig.7\subref{fig.local2_R} demonstrates that our hyper-RNN decoder has a greater advantage than the conventional BP decoder for transmission over the Rayleigh channel, where the 6-iteration hyper-RNN decoder outperforms the 150-iteration BP decoder.
In addition, {Fig.\ref{fig.finiteblock} demonstrates the BLER comparisons among the $\mathcal{O}(n^{-2})$ meta-converse Polyanskyi-Poor-Verd$\rm{\grave{u}}$ (PPV) upper bound \cite{7589108Erseghe}, the hyper-RNN decoder, and the conventional BP decoder together with the SCL decoder without CRC at the code rate of $R=1/2$ for transmission over an AWGN channel at $P_\textrm{out}=0.078$mW. The QPSK modulation is assumed.} The PPV upper bound is proposed for the finite block-length regime, which bounds the achievable BLER as a function of SNR. {However, the PPV bound has to consider the impact of energy transfer in the IDEN scenarios. Therefore, we derived the PPV bound under the target energy constraint based on the bound obtained from \cite{7589108Erseghe} under pure communication conditions. The PPV bound reflects a direct mapping relationship between the BLER and the SNR, where $SNR= 10lg(P_\textrm{tr}/{P_\textrm{n}})$. However, for a given $P_{tar}$ in the IDEN system, we can obtain the corresponding power splitting factor $\rho$ through training, thus we have $SNR'=10lg(\rho P_\textrm{tr}/ P_\textrm{n})=SNR+10lg\rho$ in the communication branch. Therefore, the mapping between the BLER and $SNR'$ in our IDEN system is also obtained.
We can observe that the BLER performance of our hyper-RNN decoder approaches the bound within $2.7$ dB provided by the PPV upper bound. Compared to the conventional BP decoder, the hyper-RNN decoder achieves up to $0.5$ dB gain, indicating that the DNN based learning methods are indeed effective. However, the SCL decoder still outperforms its BP based counterparts. This comparison reflects the superiority of both the polar code, as well as of our proposed hyper-RNN decoder, which achieves performance improvements over traditional BP decoders. This also confirms the benefits of the proposed SCL decoder.}

Moreover, in order to reflect on the adaptability advantage of our hyper-RNN decoder, we compare the BER performance of the proposed decoder and of the pure DNN-based decoder, when the number of training iterations $T_\textrm{train}$ is different from the test iteration number $T_\textrm{test}$. Specifically, we train these two decoders using $T_\textrm{train}=6$, while test them with the aid of $T_\textrm{test}=3,6$, respectively. In Fig.7\subref{fig.local2}, we can observe that our proposed decoder exhibits a better BER performance than its counterparts. This is attribute to the characteristic of hypernetwork. During the training process, the hypernetwork takes the number of iterations $t=1,2,\dots,6$ as its input in turn, and generates the corresponding weights for the RNN. Thus, the RNN can be trained and optimized under the condition of $t=1,2,\dots,6$, and obtain the corresponding network parameters to attain a better BER performance when facing different number of iterations in the test.
By contrast, the DNN-based decoder does not benefit from adaptability. Hence it cannot perform well for a number of test iterations $T_\textrm{test}$ that is different from the number of training iterations $T_\textrm{train}$. Because the DNN does not process the variable $t$ as its input to help adjust its trainable parameters during the training process, the DNN will be unable to generate the corresponding network parameters, when tested under different number of iterations. In Rayleigh channels, we also adopt the same training and testing methods as in the AWGN channel. In Fig.7\subref{fig.local2_R}, we can observe that the hypernetwork also works well in the Rayleigh channel, allowing the hyper-RNN decoder achieve a better adaptability than the pure DNN-aided decoder.%


\begin{table*}[h]
\centering
\footnotesize
\caption{Analysis of computational complexity }
\begin{tabular}{|l|c|c|c|c| } 
 \hline
     Algorithms & \tabincell{c}{Big-O\\ complexity} & Addition & Multiplication & Memory \\ \hline
 BP & $\mathcal{O}(TN\textrm{log}N)$ & \tabincell{c}{$2TN\textrm{log}N$\\ $\sim38400$} & 0 & 0 \\ 
 \hline
 DNN-based BP & $\mathcal{O}(TN\textrm{log}N)$ & \tabincell{c}{$2TN\textrm{log}N$\\ $\sim4608$} & \tabincell{c}{$2TN\textrm{log}N$\\ $\sim4608$} & \tabincell{c}{$2TN\textrm{log}N$\\ $\sim4608$} \\
 \hline
 Hyper-RNN & \tabincell{c}{$\mathcal{O}(k_hTN\textrm{log}N$\\$+k_h^2TL)$} & \tabincell{c}{$2TN\textrm{log}N$\\ $\sim4608$} & \tabincell{c}{$2TN\textrm{log}N+k_hTN\textrm{log}N$\\$+(L-1)k_h^2T+k_hT$\\$\sim23856$} & \tabincell{c}{$3N\textrm{log}N+k_hL$\\$\sim1176$} \\
 \hline
\end{tabular}
\label{tab.compare}
\end{table*} 

Further analysis of the hyper-RNN decoder's computational complexity  together with the conventional BP decoder and the DNN-based decoder are shown in Table \ref{tab.compare}, where $N$ denotes the code length, $T$ is the number of iterations and $k_h$ denotes the number of neurons in each hidden layer of the hypernetwork. Firstly, we can observe that the hyper-RNN has a higher Big-O complexity than its counterparts. This is mainly due to the hypernetwork, where the number of hidden layers $L$ and the number of neurons in each layer $k_h$ have substantial effects. {Moreover, the simulation results reveal that the $50$-iteration BP decoder, the $6$-iteration DNN-based BP decoder and the $6$-iteration hyper-RNN decoder exhibit comparable BER performance under an AWGN channel. Here we set $L=3$, $N=64$ and $k_h=8$. It is observed that both the DNN-based BP and the hyper-RNN significantly reduce the the number of addition operations compared to the conventional BP due to their superiority in decoding. However, this comes at the cost of increasing the number of multiplication operations and memory overhead. The hyper-RNN requires more multiplication operations than the DNN-based BP because of the existence of an extra hypernetwork. Nevertheless, the hyper-RNN reduces the memory overhead as a benefit of the weight sharing property. In addition, the hyper-RNN decoder's adaptability is improved, thus the complexity is acceptable.} 


\subsection{Global Performance Comparison}
\begin{figure}[h]
	\centering  
	\subfloat[AWGN]{
	\label{fig.globalA}
		\includegraphics[width=2.9in]{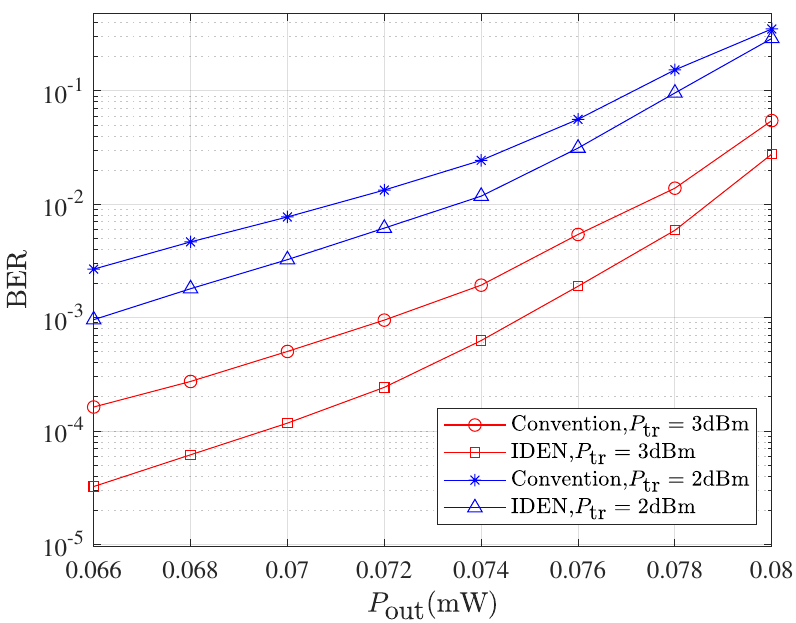}}\\
	\subfloat[Rayleigh]{
	\label{fig.global}
	\includegraphics[width=2.9in]{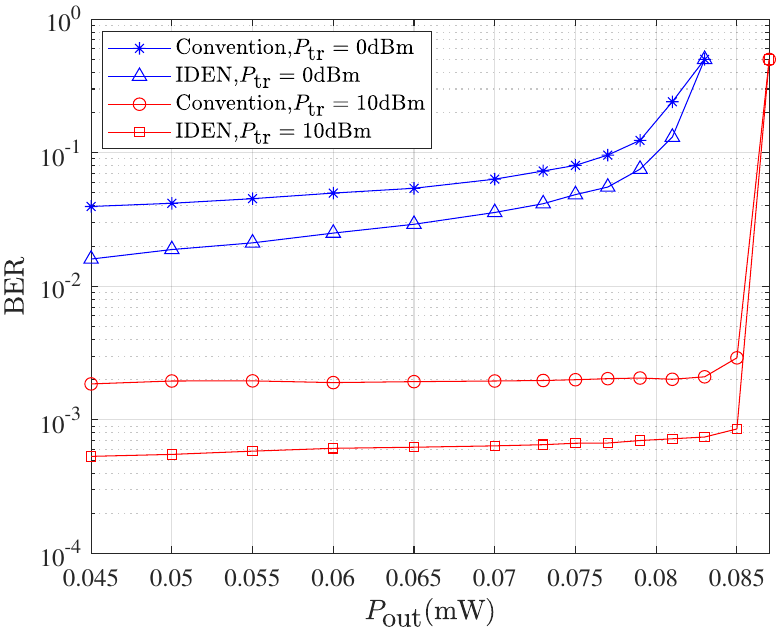}}
	\caption{The global performance comparisons between the proposed IDEN and the conventional counterparts (a) Performance over AWGN channels for iter = 6 decoding iterations. (b) Performance over Rayleigh channels for iter = 6 decoding iterations}
	\label{fig.globalAR}
	\vspace{-0.2 cm}
\end{figure}

Fig.\ref{fig.globalAR} demonstrates the global performance comparisons of the end-to-end DNN-aided IDEN system and of the BP-based conventional IDEN system for transmission over both an AWGN channel and a Rayleigh channel. The transmit power is set to {$P_{\textrm{tr}}=2$ dBm and $P_{\textrm{tr}}=3$ dBm} for the AWGN channel and $P_{\textrm{tr}}=0$ dBm and $P_{\textrm{tr}}=10$ dBm for the Rayleigh channel. {The number of decoding iterations is set to $6$ for both the AWGN channel and the Rayleigh channel}. Observe that our proposed IDEN system is superior to the conventional one under both 
the AWGN channel and Rayleigh channel in terms of its WET and WIT performance. {The proposed IDEN system achieves a lower BER for the same energy requirements, showing its optimality. For example, when $P_{\textrm{tr}}=10$ dBm in a Rayleigh channel, the proposed IDEN system reduces the BER from $1.8\times 10^{-3}$ to $5.3\times 10^{-4}$ at the harvested energy of $P_{\textrm{out}}=0.045$ mW.} This is mainly due to the 
effect of each NN-based module in the system, where the BNN-aided polar encoder determines the appropriate frozen bit positions, the DNN based AE-mapper generates the constellations which satisfies the WET and WIT requirements simultaneously, and the hyper-RNN decoder improves the decoding capability and enhances the adaptability. The end-to-end joint training allows the system to adjust the training parameters in a timely manner according to the feedback and hence achieves better WIT and WET performance.
Moreover, it is worth noting that although the modulation order is set to $4$, the learned constellation may not necessarily be identical to 4QAM. This is because the AE-mapper only learns a mapping relationship from binary bit sequences to modulated symbols, which is influenced by practical communication and energy requirements. A more detailed analysis of the AE-learned constellations will be discussed in the follow. 

\subsection{Constellations}
\begin{figure}[t]
	\centering  

	\subfloat[]{
	\label{fig.original}		\includegraphics[width=1.7in]{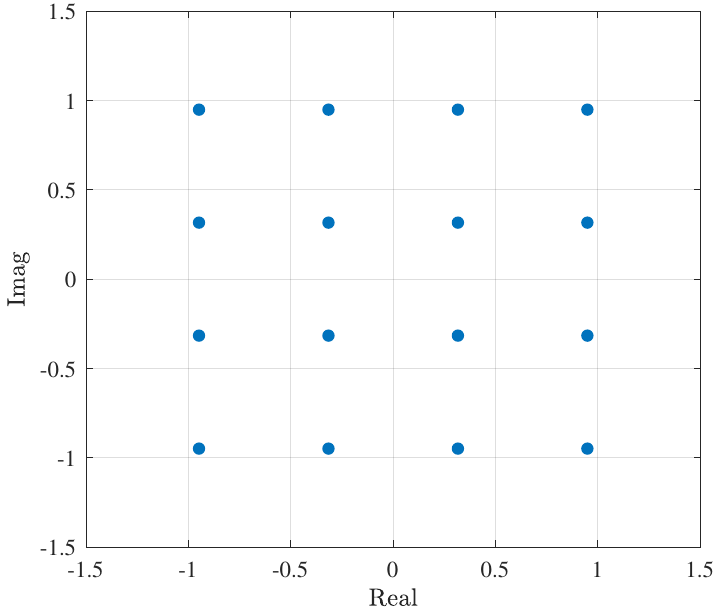}}
	\subfloat[]{
	\label{fig.M16_1}		\includegraphics[width=1.7in]{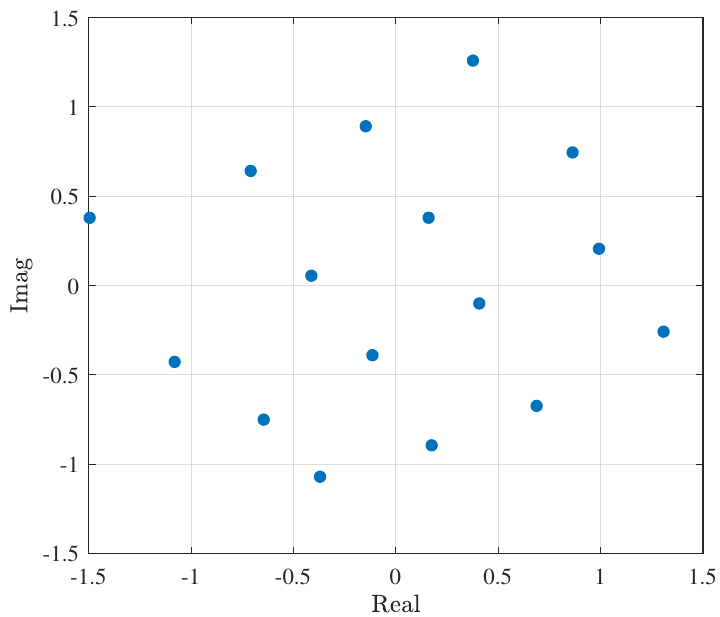}}\\
	\subfloat[]{
	\label{fig.M16_2} 
	\includegraphics[width=1.7in]{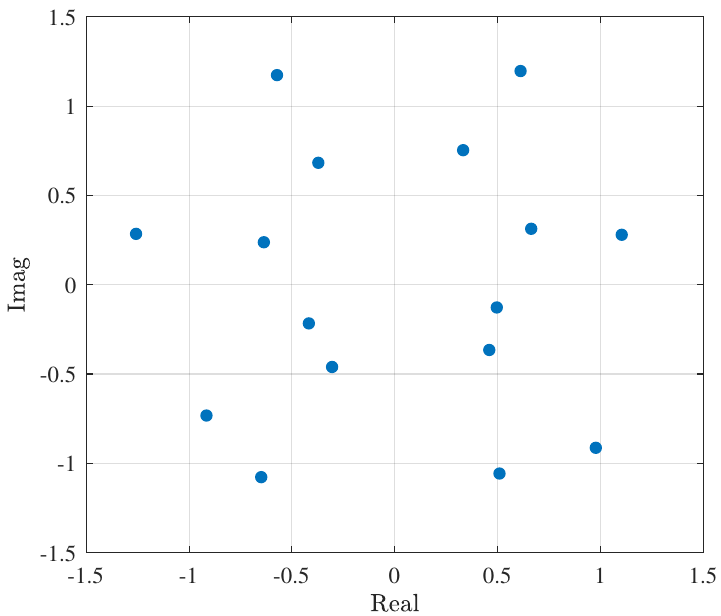}}
        \subfloat[]{
	\label{fig.M16_3} 
	\includegraphics[width=1.7in]{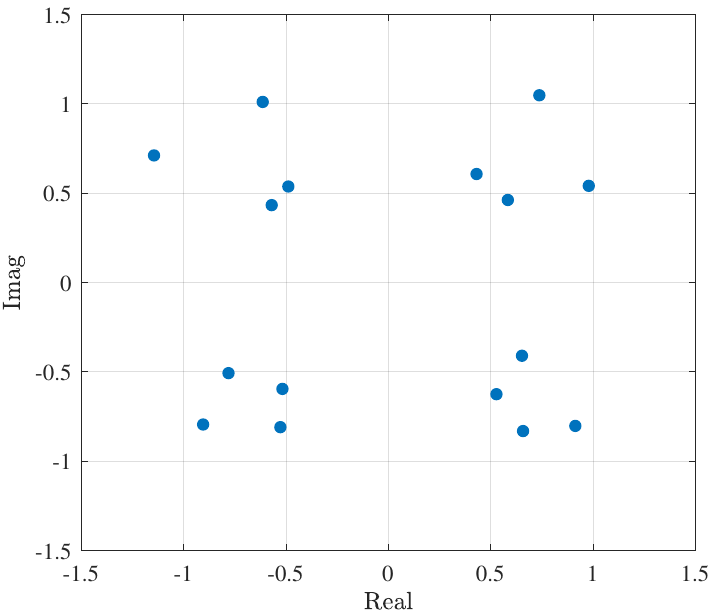}}
 
	\caption{Constellations of (a) 16QAM and AE in different energy demands: (b) pure WIT, (c) WET-aware, $P_{\textrm{out}}=0.048mW$, (d) WET-aware, $P_{\textrm{out}}=0.074mW$ .}
	\label{fig.constellation}
	\vspace{-0.3 cm}
\end{figure}

To further exemplify the effect of the trade-off between WIT and WET on the constellations output by the AE-mapper in the system, we present the constellations learned along with $M=16$ for different energy requirements in Fig.\ref{fig.constellation}. Firstly, we can observe that in a pure WIT scenario without energy demand, the distribution of constellation points learned by the AE-mapper is relatively uniform and forms a grid, which is similar to 16-QAM. It only aims to minimize the BER. However, when considering WET, the constellation geometry changes significantly, as shown in Fig.10\subref{fig.M16_2} and Fig.10\subref{fig.M16_3}. The distribution of constellation points is approximately of circular shape, and compared to Fig.10\subref{fig.M16_1}, the low-energy points move away from the origin as the energy demand increases, thus carrying higher energy. Meanwhile, the distance between constellation points is shortened, leading to an increase in BER. This reflects the trade-off between WET and WIT.

\begin{figure}[h]
  \centering
  \includegraphics[width=3.1in]{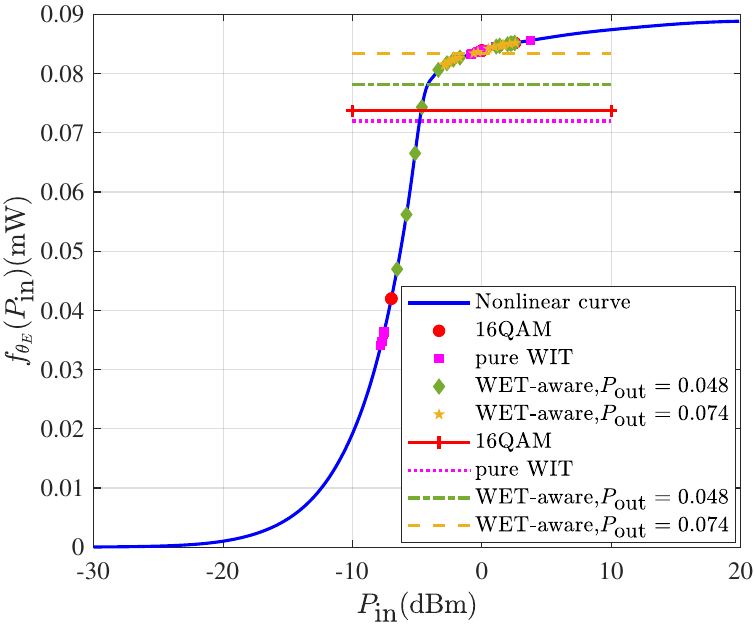}\\
  \caption{The mapping of constellations on the nonlinear curve of EH.}\label{fig.curve_16_1}
  \vspace{-0.4 cm}
\end{figure}

{Furthermore, in order to demonstrate the energy improvements of the constellations from a numerical perspective, we map the constellation points (labeled as filled markers) onto the nonlinear input-output curve of the EH based on the energy they carry. Then we calculate the average output energy (labeled as horizontal lines) in Fig.\ref{fig.curve_16_1}, thus we can observe the energy changes in the constellations more intuitively. In a pure WIT scenario, many constellation points output by the AE-mapper carry low energy, since there is no energy demand. However, as the energy demand increases, the modulation is reformed and the output constellation is adjusted to achieve higher harvested energy at the receiver under nonlinear conditions. It can be observed that more constellation points are moving along the curve towards the high-energy region, and correspondingly, the average output energy is also significantly improved. Indeed, it becomes even higher than that of 16-QAM. These results indicate that the energy of the learned constellations increases with the energy demand of the receiver, underlining the advantages of the learning-based modulation design over its conventional counterparts in IDEN systems.}

\subsection{Training Convergence}

\begin{figure}[h]
  \centering
  \includegraphics[width=3in]{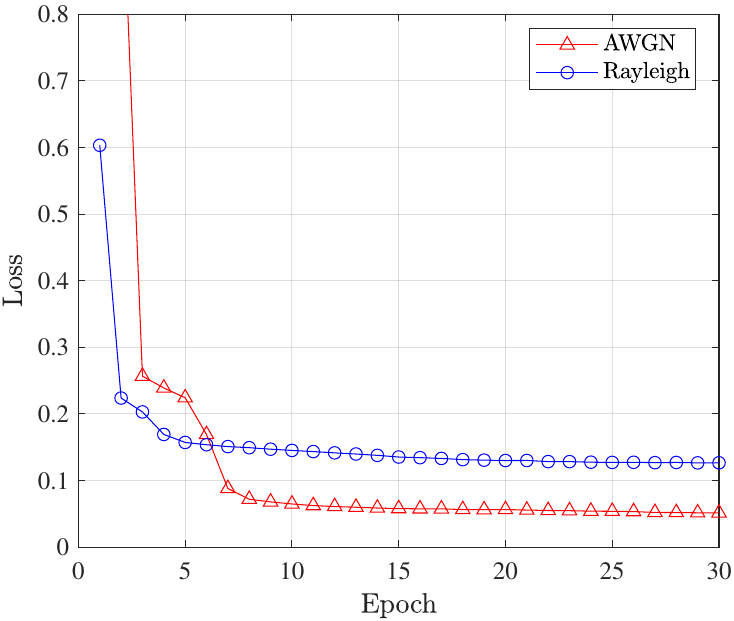}\\
  \caption{Convergence trend of loss during training over an AWGN channel with $P_{\textrm{tr}}=5$ dBm and over a Rayleigh channel with $P_{\textrm{tr}}=0$ dBm.}\label{fig.loss}
\end{figure}
Fig.\ref{fig.loss} demonstrates the trends of the loss curves during training for transmission over both an AWGN channel and a Rayleigh channel. Observe that the two curves exhibit similar trends. The loss value is relatively high at the beginning of training, because the network is just getting trained and the network parameters have not been adjusted to the appropriate values. After a few epochs of training, the loss curves rapidly decline, which indicates that the network parameters have been improved. This results in a significant improvement of the IDEN performance of the system. During the subsequent training, the loss curves remain stable, which indicates that the training of the network has reached convergence and that the network parameters have been optimized.

\section{Conclusions}
In this paper, we designed a deep learning-aided end-to-end polar coded IDEN system, where all the modules are replaced by DNNs and trained jointly in an end-to-end manner for maximizing the communication performance, while satisfying the energy requirements in the IDEN scenarios considered. We emphasized the impact of polar encoding on the WET performance improvement by constructing the BNN-aided polar encoder which generates trainable frozen bit positions to adjust to time-variant channels. We also applied hypernetworks to the hyper-RNN decoder for better adaptability.
The simulation results collected in both AWGN and Rayleigh channels quantify the WIT and WET performance improvements attained by the BNN-aided polar encoder and the BER performance of the hyper-RNN decoder. Moreover, the global performance of our proposed system is significantly superior to that of its BP-based traditional counterparts in terms of both WIT and WET. {However, for the cases where the performance of our proposed system is inferior to that of the SCL-based counterparts, further research is needed to seek for better designs to surpass the performance of SCL-based systems.}

\bibliographystyle{ieeetr}	
\bibliography{mm}

\begin{thebibliography}{10}

\bibitem{hu2018integrated}
J.~Hu, K.~Yang, G.~Wen, and L.~Hanzo, ``Integrated data and energy communication network: A comprehensive survey,'' {\em IEEE Communications Surveys \& Tutorials}, vol.~20, no.~4, pp.~3169--3219, 2018.

\bibitem{clerckx2022foundations1}
B.~Clerckx, J.~Kim, K.~W. Choi, and D.~I. Kim, ``Foundations of wireless information and power transfer: Theory, prototypes, and experiments,'' {\em Proceedings of the IEEE}, vol.~110, no.~1, pp.~8--30, 2022.

\bibitem{clerckx2021wireless1}
B.~Clerckx, K.~Huang, L.~R. Varshney, S.~Ulukus, and M.-S. Alouini, ``Wireless power transfer for future networks: Signal processing, machine learning, computing, and sensing,'' {\em IEEE Journal of Selected Topics in Signal Processing}, vol.~15, no.~5, pp.~1060--1094, 2021.

\bibitem{clerckx2018fundamentals}
B.~Clerckx, R.~Zhang, R.~Schober, D.~W.~K. Ng, D.~I. Kim, and H.~V. Poor, ``Fundamentals of wireless information and power transfer: From {RF} energy harvester models to signal and system designs,'' {\em IEEE Journal on Selected Areas in Communications}, vol.~37, no.~1, pp.~4--33, 2019.

\bibitem{varasteh2020learning}
M.~Varasteh, J.~Hoydis, and B.~Clerckx, ``Learning to communicate and energize: Modulation, coding, and multiple access designs for wireless information-power transmission,'' {\em IEEE Transactions on Communications}, vol.~68, no.~11, pp.~6822--6839, 2020.

\bibitem{xiang2023polar}
L.~Xiang, J.~Cui, J.~Hu, K.~Yang, and L.~Hanzo, ``Polar coded integrated data and energy networking: A deep neural network assisted end-to-end design,'' {\em IEEE Transactions on Vehicular Technology}, pp.~1--6, 2023.

\bibitem{fouladgar2014constrained}
A.~M. Fouladgar, O.~Simeone, and E.~Erkip, ``Constrained codes for joint energy and information transfer,'' {\em IEEE Transactions on Communications}, vol.~62, no.~6, pp.~2121--2131, 2014.

\bibitem{tandon2016subblock}
A.~Tandon, M.~Motani, and L.~R. Varshney, ``Subblock-constrained codes for real-time simultaneous energy and information transfer,'' {\em IEEE Transactions on Information Theory}, vol.~62, no.~7, pp.~4212--4227, 2016.

\bibitem{basu2019polar}
S.~Basu and L.~R. Varshney, ``Polar codes for simultaneous information and energy transmission,'' in {\em 2019 IEEE 20th International Workshop on Signal Processing Advances in Wireless Communications (SPAWC)}, pp.~1--5, IEEE, 2019.

\bibitem{dabirnia2016code}
M.~Dabirnia and T.~M. Duman, ``On code design for joint energy and information transfer,'' {\em IEEE Transactions on Communications}, vol.~64, no.~6, pp.~2677--2688, 2016.

\bibitem{kim2021wireless}
J.~Kim and B.~Clerckx, ``Wireless information and power transfer for {IoT}: Pulse position modulation, integrated receiver, and experimental validation,'' {\em IEEE Internet of Things Journal}, vol.~9, no.~14, pp.~12378--12394, 2022.

\bibitem{zeng2017communications}
Y.~Zeng, B.~Clerckx, and R.~Zhang, ``Communications and signals design for wireless power transmission,'' {\em IEEE Transactions on Communications}, vol.~65, no.~5, pp.~2264--2290, 2017.

\bibitem{clerckx2016waveform}
B.~Clerckx and E.~Bayguzina, ``Waveform design for wireless power transfer,'' {\em IEEE Transactions on Signal Processing}, vol.~64, no.~23, pp.~6313--6328, 2016.

\bibitem{ku2017joint}
M.-L. Ku, Y.~Han, B.~Wang, and K.~R. Liu, ``Joint power waveforming and beamforming for wireless power transfer,'' {\em IEEE Transactions on Signal Processing}, vol.~65, no.~24, pp.~6409--6422, 2017.

\bibitem{collado2014optimal}
A.~Collado and A.~Georgiadis, ``Optimal waveforms for efficient wireless power transmission,'' {\em IEEE Microwave and Wireless Components Letters}, vol.~24, no.~5, pp.~354--356, 2014.

\bibitem{huang2017waveform}
Y.~Huang and B.~Clerckx, ``Waveform design for wireless power transfer with limited feedback,'' {\em IEEE Transactions on Wireless Communications}, vol.~17, no.~1, pp.~415--429, 2017.

\bibitem{varasteh2020oncapacity}
M.~Varasteh, B.~Rassouli, and B.~Clerckx, ``On capacity-achieving distributions for complex awgn channels under nonlinear power constraints and their applications to {SWIPT},'' {\em IEEE Transactions on Information Theory}, vol.~66, no.~10, pp.~6488--6508, 2020.

\bibitem{clerckx2018beneficial0}
B.~Clerckx and J.~Kim, ``On the beneficial roles of fading and transmit diversity in wireless power transfer with nonlinear energy harvesting,'' {\em IEEE Transactions on Wireless Communications}, vol.~17, no.~11, pp.~7731--7743, 2018.

\bibitem{clerckx2017wireless11}
B.~Clerckx, ``Wireless information and power transfer: Nonlinearity, waveform design, and rate-energy tradeoff,'' {\em IEEE Transactions on Signal Processing}, vol.~66, no.~4, pp.~847--862, 2017.

\bibitem{varasteh2019learning2}
M.~Varasteh, J.~Hoydis, and B.~Clerckx, ``Learning modulation design for {SWIPT} with nonlinear energy harvester: Large and small signal power regimes,'' in {\em 2019 IEEE 20th International Workshop on Signal Processing Advances in Wireless Communications (SPAWC)}, pp.~1--5, IEEE, 2019.

\bibitem{9583222Hameed}
I.~Hameed, P.~V. Tuan, and I.~Koo, ``Deep learning–based energy beamforming with transmit power control in wireless powered communication networks,'' {\em IEEE Access}, vol.~9, pp.~142795--142803, 2021.

\bibitem{8469031Kang}
J.-M. Kang, C.-J. Chun, and I.-M. Kim, ``Deep-learning-based channel estimation for wireless energy transfer,'' {\em IEEE Communications Letters}, vol.~22, no.~11, pp.~2310--2313, 2018.

\bibitem{shanin2020rate}
N.~Shanin, L.~Cottatellucci, and R.~Schober, ``Rate-power region of {SWIPT} systems employing nonlinear energy harvester circuits with memory,'' in {\em ICC 2020-2020 IEEE International Conference on Communications (ICC)}, pp.~1--7, IEEE, 2020.

\bibitem{arikan2009channel}
E.~Arikan, ``Channel polarization: A method for constructing capacity-achieving codes for symmetric binary-input memoryless channels,'' {\em IEEE Transactions on information Theory}, vol.~55, no.~7, pp.~3051--3073, 2009.

\bibitem{9097207Xu}
W.~Xu, X.~Tan, Y.~Be’ery, Y.-L. Ueng, Y.~Huang, X.~You, and C.~Zhang, ``Deep learning-aided belief propagation decoder for polar codes,'' {\em IEEE Journal on Emerging and Selected Topics in Circuits and Systems}, vol.~10, no.~2, pp.~189--203, 2020.

\bibitem{8863406Liu}
X.~Liu, S.~Wu, Y.~Wang, N.~Zhang, J.~Jiao, and Q.~Zhang, ``Exploiting error-correction-crc for polar {SCL} decoding: A deep learning-based approach,'' {\em IEEE Transactions on Cognitive Communications and Networking}, vol.~6, no.~2, pp.~817--828, 2020.

\bibitem{8109997}
W.~Xu, Z.~Wu, Y.-L. Ueng, X.~You, and C.~Zhang, ``Improved polar decoder based on deep learning,'' in {\em 2017 IEEE International Workshop on Signal Processing Systems (SiPS)}, pp.~1--6, 2017.

\bibitem{teng2019low}
C.-F. Teng, C.-H.~D. Wu, A.~K.-S. Ho, and A.-Y.~A. Wu, ``Low-complexity recurrent neural network-based polar decoder with weight quantization mechanism,'' in {\em ICASSP 2019-2019 IEEE International Conference on Acoustics, Speech and Signal Processing (ICASSP)}, pp.~1413--1417, IEEE, 2019.

\bibitem{ebada2019deep}
M.~Ebada, S.~Cammerer, A.~Elkelesh, and S.~ten Brink, ``Deep learning-based polar code design,'' in {\em 2019 57th Annual Allerton Conference on Communication, Control, and Computing (Allerton)}, pp.~177--183, IEEE, 2019.

\bibitem{8962344Bioglio}
V.~Bioglio, C.~Condo, and I.~Land, ``Design of polar codes in {5G} new radio,'' {\em IEEE Communications Surveys \& Tutorials}, vol.~23, no.~1, pp.~29--40, 2021.

\bibitem{mori2009performance}
R.~Mori and T.~Tanaka, ``Performance of polar codes with the construction using density evolution,'' {\em IEEE Communications Letters}, vol.~13, no.~7, pp.~519--521, 2009.

\bibitem{3rd2021technical}
3rd Generation Partnership~Project, ``Technical specification group radio access network; nr.; multiplexing and channel coding (release 16), 3gpp ts 38.212 v16. 5.0 (2021-03),'' 2021.

\bibitem{babar2019polar}
Z.~Babar, Z.~B.~K. Egilmez, L.~Xiang, D.~Chandra, R.~G. Maunder, S.~X. Ng, and L.~Hanzo, ``Polar codes and their quantum-domain counterparts,'' {\em IEEE Communications Surveys \& Tutorials}, vol.~22, no.~1, pp.~123--155, 2019.

\bibitem{Egilmez8936409}
Z.~B.~K. Egilmez, L.~Xiang, R.~G. Maunder, and L.~Hanzo, ``The development, operation and performance of the {5G} polar codes,'' {\em IEEE Communications Surveys \& Tutorials}, vol.~22, no.~1, pp.~96--122, 2019.

\bibitem{cammerer2020trainable}
S.~Cammerer, F.~A. Aoudia, S.~D{\"o}rner, M.~Stark, J.~Hoydis, and S.~Ten~Brink, ``Trainable communication systems: Concepts and prototype,'' {\em IEEE Transactions on Communications}, vol.~68, no.~9, pp.~5489--5503, 2020.

\bibitem{liu2013wireless}
L.~Liu, R.~Zhang, and K.-C. Chua, ``Wireless information and power transfer: A dynamic power splitting approach,'' {\em IEEE Transactions on Communications}, vol.~61, no.~9, pp.~3990--4001, 2013.

\bibitem{4178493}
R.~A. Shafik, M.~S. Rahman, and A.~R. Islam, ``On the extended relationships among evm, ber and snr as performance metrics,'' in {\em 2006 International Conference on Electrical and Computer Engineering}, pp.~408--411, 2006.

\bibitem{arikan2010graph}
E.~Ar{\i}kan, ``Polar codes: A pipelined implementation,'' in {\em Proc. 4th ISBC}, vol.~2010, pp.~11--14, 2010.

\bibitem{goodfellow2016deep}
I.~Goodfellow, Y.~Bengio, and A.~Courville, {\em Deep learning}.
\newblock MIT press, 2016.

\bibitem{7589108Erseghe}
T.~Erseghe, ``Coding in the finite-blocklength regime: Bounds based on laplace integrals and their asymptotic approximations,'' {\em IEEE Transactions on Information Theory}, vol.~62, no.~12, pp.~6854--6883, 2016.

\end{thebibliography}

\end{document}